\documentclass[12pt]{article}
\usepackage{epsf}
\usepackage{graphicx}
\usepackage{amssymb}
\usepackage{epstopdf}
\usepackage{amsmath}
\usepackage{amssymb}
\usepackage{epsfig}
\usepackage{cite}
\usepackage{color,colordvi}
\usepackage{graphicx}
\usepackage{pdfpages}
\usepackage{appendix}
\graphicspath{ {./} }
\usepackage[left=2.5cm,bottom=3cm,right=2.5cm,top=3cm]{geometry} 

\usepackage[latin,english]{babel}
\usepackage{amsmath}
\usepackage{amssymb}
\usepackage{graphicx}
\usepackage{mathtools}
\usepackage{tikz} 
\usetikzlibrary{patterns}
\usetikzlibrary{decorations.markings,decorations.pathmorphing}
\usetikzlibrary{calc}
\tikzstyle{singularity}=[line width=0.6,decorate,
                         decoration={zigzag,amplitude=2,segment length=6.17}]
\usetikzlibrary{shapes.misc}
\colorlet{mydarkred}{red!50!black}
\usepackage{subcaption}

\colorlet{mydarkred}{red!50!black}
\colorlet{myblue}{blue!50!black}
\colorlet{mylightblue}{myblue!6}
\colorlet{myred}{red!80!black}
\colorlet{mygreen}{green!80!black}
\colorlet{mydarkred}{red!50!black}
\colorlet{myblue}{blue!50!black}
\colorlet{mylightblue}{myblue!6}
\colorlet{mypurple}{blue!40!red!80!black}
\colorlet{mydarkpurple}{blue!40!red!50!black}
\colorlet{mylightpurple}{mydarkpurple!80!red!6}
\colorlet{myorange}{orange!40!yellow!95!black}
\definecolor{mypink}{HTML}{FDA4BA}
\tikzset{cross/.style={cross out,line width=2pt, rotate=25, draw=mydarkred, minimum size=2*(#1-\pgflinewidth), inner sep=0pt, outer sep=0pt},
cross/.default={1pt}}
\usetikzlibrary{patterns}
\usetikzlibrary{decorations.markings,decorations.pathmorphing}
\usetikzlibrary{calc}
\tikzstyle{singularity}=[line width=0.6,decorate,mydarkred,
                         decoration={zigzag,amplitude=3,segment length=8.25}]
\usetikzlibrary{shapes.misc}

\begin{document}
\vspace{0.01cm}
\begin{center}
{\Large\bf  Duality and Black Hole Evaporation} 

\end{center}

\vspace{0.1cm}

\begin{center}

{\bf Cesar Gomez}$^{a}$\footnote{cesar.gomez@uam.es}

\vspace{.6truecm}

{\em $^a$
Instituto de F\'{\i}sica Te\'orica UAM-CSIC\\
Universidad Aut\'onoma de Madrid,
Cantoblanco, 28049 Madrid, Spain}\\

\end{center}


\begin{abstract}
\noindent  
 
{We construct a model that describes the quantum black hole evaporation unitarily in a Hilbert space of infinite dimension. This construction generalizes Page's finite dimensional approach to infinite dimensions. The basic ingredient is the Murray-von Neumann coupling for finite type II factors. This coupling measures, at each time of the evaporation, the relative continuous dimension of the radiation and the black hole subspaces. The unitary transformation, implementing the quantum evaporation and thus determining the time dependence of the coupling, is identified with the dual modular automorphism. In the appendix we sketch, using von Neumann construction of infinite tensor products of EPR pairs of q-bits, some q-bit holographic correspondences as well as an algebraic definition of ER bridges.

}

\end{abstract}

\thispagestyle{empty}
\clearpage
\tableofcontents
\newpage

%
\section{Introduction}
Since its original enunciation by Hawking \cite{Hawking1}, the information paradox has been one of the main ways of identifying the conceptual problems that confront us in the construction of a quantum theory of gravity. If, in the process of black hole formation and evaporation, the initial conditions are defined by a pure state then {\it unitarity} implies that the von Neumann entropy at the end of the evaporation process should be also zero. However, in the intermediate moments of the process, the {\it external observer} can define a non-zero von Neumann entropy. This entropy is the entropy measuring, along the process, the quantum entanglement between the asymptotically observable radiated quanta and the black hole interior. The information problem lies in discovering how, as required by unitarity, this quantum entanglement entropy becomes zero at the end of the evaporation. 

In the {\it semiclassical} description of the evaporation process developed by Hawking \cite{Hawking2} any radiated quanta is maximally entangled to the corresponding Hawking partner inside the black hole. Consequently the entanglement entropy of the radiation will increase in a way proportional to the number of emitted quanta.
 This fact led Hawking to suggest that in the process of formation and evaporation of a black hole the von Neumann entropy increases or in other words that information is lost. 

In slightly more precise terms, the paradox appears at the moment in which the von Neumann entanglement entropy of the Hawking radiation, let us say $S_{rad}^{Hawking}(s)$, is greater than the Bekenstein-Hawking thermodynamic entropy \cite{Bek1} of the black hole $S^{Th}(s)$. What is paradoxical is that, in the semiclassical approach, this problem occurs at moments of evaporation where the semiclassical approximation is a priori perfectly valid \footnote {Roughly at the moment in which a number of quanta of the order of half the value of the original entropy of the black hole have been emitted.}. The natural solution to this paradox is to assume that in the gravitational case the definition of the von Neumann entropy must be modified in such a way that it respects {\it unitarity} i.e $S_{rad}\leq S^{Th}$ along the full evaporation process \footnote{For a review of the information paradox see \cite{Mathur}.}.

Recently a very interesting modification of the definition of von Neumann entropy in the presence of gravity, preserving unitarity, has been proposed combining two key ideas: Bekenstein's definition of {\it generalized entropy} \cite{Bek2} and the holographic RT \cite{RT,HRT} version of the entanglement entropy. This leads to the so called QES formalism \cite{EW,QES1,QES2,QES3,QES4} \footnote{For a review of the application of QES formalism to the information paradox see \cite{Malda}.}. Intuitively the modified von Neumann entropy of the radiation is obtained substracting to the naiv quantum entanglement entropy, that goes as $O(n)$ for $n$ the number of radiated quanta, an {\it island} contribution $I(n)$ of order $2n-N$ for $N$ the black hole entropy at the moment of formation and for $n\geq \frac{N}{2}$. The very non trivial fact is that a continuous and time dependent version of $I(n)$ can be represented as the area of the {\it quantum extremal surface} $\frac{Area(QES)}{4G_N}$ \footnote{In this note we will not discuss further the QES prescription. See \cite{Gomez1} for some comments.}. 

A more direct and also more radical solution to the paradox can be designed on the basis of Page's analysis \cite{Page} if we assume that the evaporation process can be described, {\it by the external observer}, not only in a {\it unitary} way but also in a {\it finite-dimensional} Hilbert space $H$ of some finite dimension $N$. Under these conditions, the evaporation process can be described by a single parameter defined, at each instant of evaporation, as the quotient 
\begin{equation}
d_P(s)=\frac{dH_{rad}(s)}{dH_{BH}(s)}
\end{equation}
between the dimensions of the finite dimensional Hilbert spaces of the black hole $dH_{BH}(s)$ and the radiation $dH_{rad}(s)$. 


The full evaporation is described by a {\it growing function} $d_P(s)$ ( see Figure1) that will start at the moment $s=0$ of black hole formation with a value $O(\frac{1}{N})$ and will end with a value $O(N)$ \footnote{This growing function reflects the increase of the number of radiated quanta as well as the decrease in the black hole size.}.
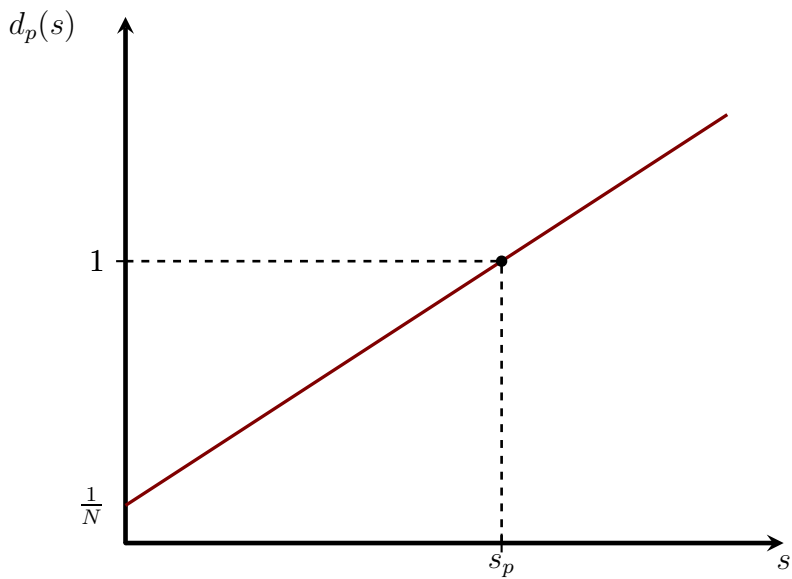
\begin{figure}[htb!]
    \begin{center}
        \begin{tikzpicture}[scale=2.5]
        \draw[-stealth,thick,line width=1.75](0,0)--(3.5,0)node[anchor=north]{$s$};
        \draw[-stealth,thick,line width=1.75](0,0)--(0,2.8)node[anchor=east]{};
        \draw[thick,line width=1,dashed](0,1.5)--(2,1.5);
        \draw[thick,line width=1,dashed](2,0)--(2,1.5);
        \draw[thick]  (-0.05,1.5)circle(0pt) node[anchor=east ] {$1$};
        \draw[thick,line width=1.25,mydarkred](3.2,2.28)--(0,.2);
        \filldraw(0,0) circle(.25pt);
        \draw[thick]  (-.2,3-.4)circle(0pt) node[anchor=south east ] {$d_p(s)$};
        \draw[thick]   (2,-0.05) --(2,0.05);
        \draw[thick]  (2,0)circle(0pt) node[anchor=north ] {$s_p$};
        \draw[thick]   (0,1.5) --(-0.05,1.5);
        \draw[thick]  (-0.05,1.5)circle(0pt) node[anchor=east ] {$1$};
        \draw[thick]  (-0.05,0.2)circle(0pt) node[anchor=east ] {$\frac{1}{N}$};
        \filldraw[thick] (2,1.5) circle (.7pt) ;
        \end{tikzpicture}
    \end{center}
    \caption{This figure represents a generic uniform process of evaporation of a black hole with initial entropy $\log N$. The concrete rate of evaporation at each time which is encoded in the $s$ dependence of $d_P(s)$ is not represented in this simple cartoon. } \label{fig:1} 
\end{figure}

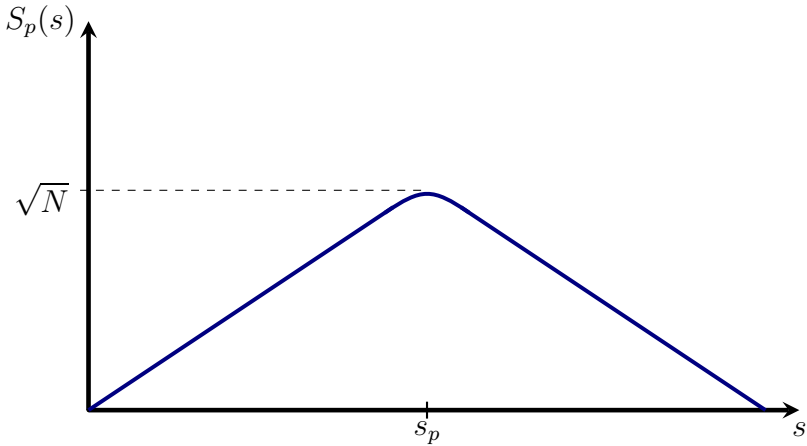
\begin{figure}
    \begin{center}
        \begin{tikzpicture}[scale=2.25]
        \draw[-stealth,thick,line width=1.75](0,0)--(4.2,0)node[anchor=north]{$s$};
        \draw[-stealth,thick,line width=1.75](0,0)--(0,2.3)node[anchor=east]{$S_p(s)$};
        \filldraw(0,0) circle(.25pt);
        \draw[thick]   (2,-0.05) --(2,0.05);
        \draw[thick]  (2,0)circle(0pt) node[anchor=north ] {$s_p$};
        \draw[myblue,line width=1.5]  (0,0)--(1.8,1.2) node[anchor=north ] {};
        \draw[myblue,line width=1.5]  (4,0)--(2.2,1.2) node[anchor=north ] {};
        \draw[myblue,line width=1.5] (1.75,1.166)..controls+(.22,.15) and +(-.22,.15)..(2.25,1.167);
        \draw[thick]  (-0.05,1.25)circle(0pt) node[anchor=east ] {$\sqrt{N}$}; 
        \draw[dashed]    (-0.05,1.3)--(2,1.3);
    \end{tikzpicture}
    \end{center}
    \caption{Typical Page curve representing $S^{vN}(s)$ in the finite dimensional Hilbert space approach.}\label{fig:2} 
\end{figure}

In these conditions the quantum evaporation process is defined by: i) a generic continuous map $s\rightarrow \psi(s)\in H$ defining the full quantum state at time $s$ and ii) a growing function $d_P(s)$. Obviously the corresponding vN entropy $S^{vN}(s)=S_{rad}(s)=S_{BH}(s)$ satisfies the {\it constraint} $S^{vN}(s)\leq S^{Th}(s)=: \log dH_{BH}(s)$ \footnote{Here by $S^{vN}(s)$ we mean the von Neumann entropy associated to the quantum state $\psi(s)$ and the factorization of the finite dimensional Hilbert space $H=H_{rad}(s)\otimes H_{BH}(s)$. This entropy is defined as $S_{rad}(s)=-tr_{H_{rad}(s)}(\rho_{\psi(s)}\log\rho_{\psi(s)})$ for $\rho_{\psi(s)}=tr_{H_{BH}(s)}|\psi(s)\rangle\langle \psi(s)|$ and the same for $S_{BH}(s)$ exchanging the roles of $H_{BH}(s)$ and $H_{rad}(s)$. The curve depicted in Figure 2 can be interpreted as the {\it average} for random states $\psi(s)$ \cite{Page}.} ( see Figure 2). 

In this finite dimensional context the question of {\it unitarity} is a bit more subtle. Since we are assuming that $H=H_{rad}(s)\otimes H_{BH}(s)$ we can define at each evaporation time $s$ a {\it projection}, let us say $P(s)$, such that $H_{rad}(s)=P(s)H$. The unitarity of the evaporation process implies the existence of  {\it unitary operators} $V_{s^{'}}$ ( acting on $H$ ) such that
\begin{equation}\label{evapo}
V_{s^{'}} P(s) V^{+}_{s^{'}} = P(s+s^{'})
\end{equation} 
and such that $P(s+s')H=H_{rad}(s+s^{'})$ is the radiation Hilbert space at time $(s+s^{'})$ \footnote{In this finite dimensional case $P_s$ is a one parameter family of projections of $H$ i.e. bounded operators acting on $H$ and satisfying $P(s)^2=P(s)$.}.

Since in presence of {\it gravity} the Hamiltonian enters as a {\it constraint} we should be careful before interpreting the unitary operator $V_s$, used to describe unitarily the {\it evaporation} process, as defined by the Hamiltonian. 

The above form of solution is based on the postulate/dogma about the {\it finiteness} of the Hilbert space \cite{Malda}. This postulate is equivalent to say that the {\it external observer} can describe the system, as well as the process, using a finite number of q-bits equal to the Bekenstein-Hawking entropy of the black hole i.e. $\frac{A}{4G_N}$ \footnote{Note that this postulate is not equivalent to identify the black hole with a quantum system with a {\it finite number of physical degrees of freedom} as it is, for instance, the case in the black hole portrait \cite{Gia}. In this second case the Hilbert space is clearly infinite dimensional.}.  
Regardless of how we can try to justify this dogma, it seems desirable to try to find a generalization of this, finite dimensional solution of the information paradox, to the case of infinite dimensional Hilbert spaces as well as a concrete identification of the {\it unitary operator driving the quantum evaporation}. This will be the target of this note.

 \vspace{0.3 cm}
 
 {\bf The proposal} The solution to the information paradox we suggest \cite{Gomez1} is based on a natural generalization, to the infinite-dimensional case, of the two basic ingredients of the finite-dimensional solution namely, the growing function $d_P(s)$ and the unitary operator $V_s$ controlling the evaporation dynamics.  
 
Regarding $d_P(s)$, measuring the relative size of the finite dimensional Hilbert spaces for the radiation and the black hole, we will define, as the infinite dimensional generalization, {\it the Murray- von Neumann coupling} \cite{MvN1} ( see also \cite{Jones} and \cite{Popa}). More specifically we will assume that:
 
  \vspace{0.3 cm}

{\bf P-1} {\it The full evaporation process is described by a one parameter family of type $II_1$ factors $M_B(s)$, $M_R(s)$ acting on an infinite dimensional Hilbert space $H$. These factors satisfy at any evaporation time $s$ the relation $M_B(s)=M_R(s)^{'}$ and they are associated, respectively, to the entanglement wedges of the black hole and the radiation. The infinite dimensional analog of $d_P(s)$ will be identified with the Murray-von Neumann coupling of $M_B(s)$ that we will denote}
\begin{equation}
d^{MvN}_B(s)
\end{equation}

 \vspace{0.3 cm}
 
 Under these conditions the necessary condition for solving the information paradox, that in the finite dimensional case was identified with the growth of $d_{P}(s)$, becomes the growth with $s$ of $d^{MvN}_B(s)$. 
 
 \vspace{0.3 cm}
 
 Concerning the question of {\it unitarity} we will use as our basic principle that:
 
  \vspace{0.3 cm}

{\bf P-2} {\it The unitary operator $V_s$ generating the evaporation process is determined by the dual (in Takesaki's sense \cite{Takesaki}) of the modular automorphism.}

\vspace{0.3 cm}

The information paradox is solved once we prove that $d^{MvN}_B(s)$ grows under the action of the {\it dual} automorphism generated by $V_s$. 

 \vspace{0.3 cm}

From a physics point of view the former statements {\bf P-1} and {\bf P-2} are obviously not very illuminating. In addition they require a more precise mathematical definition that we will develop in the rest of the note. In this introduction we will try, before entering into technical details, to provide a physics motivation.
 
  \vspace{0.3 cm}

{\bf Physics motivation.} Regarding {\bf P-1} the obvious questions are: Why type $II_1$ factors ? and Why to use the coupling $d^{MvN}_B(s)$ to measure the relative size of the radiation and black hole Hilbert spaces ?

In the finite dimensional case we can define finite projections $P(s)$ such that $P(s)H$ is the Hilbert space $H_{rad}(s)$ and to define the dimension $dH_{rad}(s)$ of the radiation Hilbert space at evaporation time $s$ simply as the trace  $tr(P(s))$. This trace will take any discrete value $1,2...N$. In this case $P(s)$ is a projection in the type $I_N$ factor defining the full algebra of bounded operators acting on the finite dimensional Hilbert space $H$ of dimension $N$. In order to extend these definitions to the infinite dimensional case i.e. $N=\infty$ we could think in replacing $I_N$ by $I_{\infty}$. Unfortunately this will not work since in $I_{\infty}$ we don't have a well defined trace. However as originally discovered by Murray and von Neumann in \cite{MvN1} we can define a {\it different} $N=\infty$ limit as a type $II_1$ factor \footnote{See Appendix A for a review of the original construction.}. The beauty of type $II_1$ factors is that they share some basic properties of the finite type $I_N$ factors. In particular, and for any projection, we can define a {\it generalized dimension} of $P(s)H$ taking continuous values in a finite interval $[0,1]$ \footnote{In case we normalize the trace.}. Note that the key difference between $I_{N}$ and type $II_1$ is that while in the $I_{N}$ case the spectrum of dimensions is {\it discrete} it becomes {\it continuous} in the type $II_1$ case \footnote{Note that from a physics point of view the {\it generalized dimension} defined by the type $II_1$ trace is the analog of the {\it microcanonical entropy}. Indeed if we describe a physical subsystem by a type $II_1$ factor we will define the {\it microcanonical entropy}, for some fixed value $a$ of some macroscopic observable $A$, as $\log rank(A-a1)$ for $rank(A-a1)$ the dimension of the space of solutions of $(A-a1)\Psi=0$. Type $II_1$ factors are those factors acting on an infinite dimensional Hilbert space for which the $rank(A-a1)$ is well defined and can take continuous values. This is the case even if the observable $A$ has {\it continuous} spectrum. More important, from a physics point of view, is that in a type $II_1$ algebra $M$ we can associate with any hermitian operator $A$ its spectral resolution i.e. $A=\int d(\lambda)E_{\lambda}$ with $E_{\lambda}$ projections in $M$. In this way and associated with a given operator $A$ representing some macroscopic observable we can define a {\it microcanonical} state as $\int d(\lambda)|\lambda\rangle|\bar\lambda\rangle d_M(E_{\lambda})$ for $E_{\lambda}$ the corresponding resolution of the identity and $d_M(E_{\lambda})$ the generalized dimension ( see A.5)}

Thus if we want to give meaning to the dimensions of the radiation and/or the black hole Hilbert spaces when we work with an infinite dimensional Hilbert space we need to work with type $II_1$ factors. However in this case we don't have a factorization of the Hilbert space $H=H_{rad}(s)\otimes H_{BH}(s)$ such that the type $II_1$ factors $M_B(s)$ and $M_R(s)$ are acting on $H_{BH}(s)$ and $H_{rad}(s)$, so: 

\vspace{0.3 cm}

{\it How to define in this case the analog of $d_P(s)$ that, in the finite dimensional case measures the relative size of the radiation and black hole Hilbert spaces ?} 

\vspace{0.3 cm}

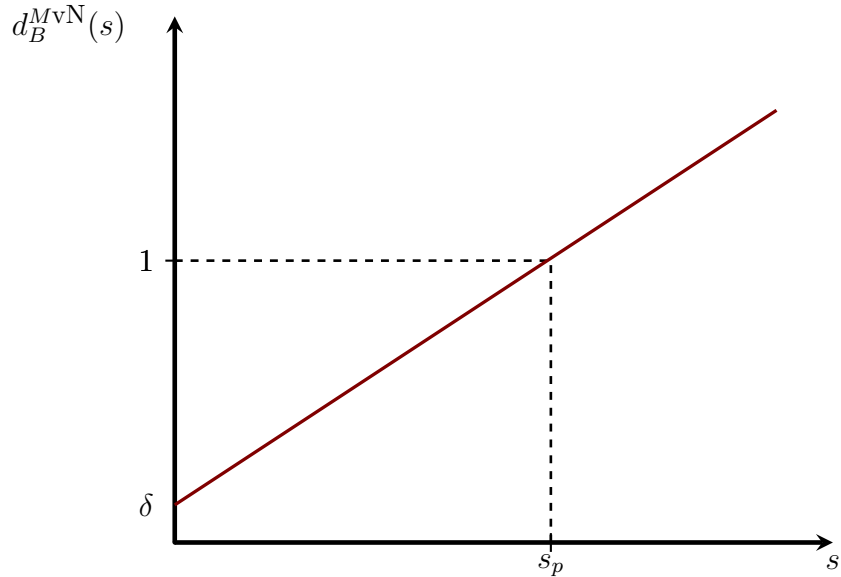
\begin{figure}
    \begin{center}
        \begin{tikzpicture}[scale=2.5]
        \draw[-stealth,thick,line width=1.75](0,0)--(3.5,0)node[anchor=north]{$s$};
        \draw[-stealth,thick,line width=1.75](0,0)--(0,2.8)node[anchor=east]{};
        \draw[thick,line width=1,dashed](0,1.5)--(2,1.5);
        \draw[thick,line width=1,dashed](2,0)--(2,1.5);
        \draw[thick]  (-0.05,1.5)circle(0pt) node[anchor=east ] {$1$};
        \draw[thick,line width=1.25,mydarkred](3.2,2.3)--(0,.2);
        \filldraw(0,0) circle(.25pt);
        \draw[thick]  (-.2,3-.4)circle(0pt) node[anchor=south east ] {$d^{M\mbox{\footnotesize vN}}_B(s)$};
        \draw[thick]   (2,-0.05) --(2,0.05);
        \draw[thick]  (2,0)circle(0pt) node[anchor=north ] {$s_p$};
        \draw[thick]   (0,1.5) --(-0.05,1.5);
        \draw[thick]  (-0.05,1.5)circle(0pt) node[anchor=east ] {$1$};
        \draw[thick]  (-0.05,0.2)circle(0pt) node[anchor=east ] {$\delta$};

        \end{tikzpicture}
    \end{center}
    \caption{Pictorical representation of a generic growing coupling along the evaporation process. For future use we have identified the infinite dimensional analog of Page's time as the value $s_P$.}\label{fig:3} 
\end{figure}

It is in order to answer this question that we need to use the Murray-von Neumann coupling $d^{MvN}_B(s)$. This coupling defines the relative {\it generalized dimension} of the subspaces $M_B(s)\psi$ and $M_R(s)\psi$ for a generic state $\psi$. Intuitively we can identify  the generalized dimensions of $M_B(s)\psi$ and $M_R(s)\psi$ as the infinite dimensional analogs of the finite dimensional quantities $dH_{BH}(s)$ and $dH_{rad}(s)$ defining $d_P(s)$. In this infinite dimensional setup the information paradox will be solved if we can prove the growth of $d^{MvN}_B(s)$ along the evaporation process ( see Figure 3).

This preliminary discussion ends the physics motivation for using type $II_1$ factors and Murray von Neumann couplings to generalize the finite $d_P(s)$ to infinite dimensions. 

 \vspace{0.3 cm}

Let us now try to motivate {\bf P-2} above. The crucial questions are: i) How to construct the unitary representation of the evaporation process and ii) What is the role played by {\it duality} in such construction. 

The lesson that we can extract from the previous discussion about the generalization to infinite dimension of the parameter $d_{P}(s)$ is that in order to describe the evaporation process we need a family of projections $P(s)$ with finite trace i.e. finite continuous dimension, and related to each other through the {\it unitary operator} representing the process of evaporation. In order to sketch the structure of the construction let us first associate with the black hole at the moment of formation an algebra ${\cal{A}}$ representing the algebra of local observables accesible to the external asymptotic observer. On the basis of Araki's result \cite{Araki} we will assume, as recently  suggested in \cite{Liu1,Liu2,Witten0,Witten1,Witten2,Witten3} that this algebra is a type $III_1$ factor. For a given state $\phi$ on ${\cal{A}}$ we can define  the corresponding GNS Hilbert space as well as the representation of the group $G$ of modular time translations in terms of the modular automorphism $\sigma_t$ \footnote{The precise definitions will be presented in section 2. Here we use a more qualitative and imprecise language with the intention of highlighting as much as possible the physical meaning of the construction.}. Physically we will like to implement the action of the group $G$ of modular time translations {\it covariantly}, in the sense of \cite{Dopli, Takesaki} i.e. we would like to define a unitary transformation $U_t$ in ${\cal{A}}$ such that $\sigma_t(a)=U_t aU_t^{+}$. For ${\cal{A}}$ a type $III_1$ the covariant implementation, in the former sense requires to define the crossed product algebra ${\cal{A}}\rtimes_{\sigma_t}{\mathbb{R}}$ \footnote{See section 2 for definitions}.

Now {\it duality} will enter into the game. Following Takesaki \cite{Takesaki} we can define the action of the {\it dual group} $\hat G$ on ${\cal{A}}\rtimes_{\sigma_t}{\mathbb{R}}$ in terms of the dual automorphism $\hat\sigma_s$. Again the {\it covariant} representation of the action of $\hat G$ requires to move into the {\it double crossed product} $({\cal{A}}\rtimes_{\sigma_t}{\mathbb{R}})\rtimes_{\hat \sigma_s} {\mathbb{R}}$. Using Takesaki duality we find that this double crossed product is isomorphic to ${\cal{A}}\otimes F_{\infty}$ with $F_{\infty}$ the type $I$ factor defined by the algebra of bounded operators acting on $L^2({\mathbb{R}})$.

This $F_{\infty}$ factor is the one interpreted as representing a quantum external {\it observer} in \cite{Witten0,Witten1,Witten2}. A lesson we can learn from this discussion is that the {\it external observer} ( in the former sense)  is {\it emerging} from duality \footnote{Strictly speaking we are not {\it adding} an observer rather the observer naturally emerges when we dualize the covariant implementation of modular transformations.}. 

Moreover using this double crossed product we can discover how the group $G$ of modular time translations is acting on ${\cal{A}}\otimes F_{\infty}$ and to identify as a  "Hamiltonian" the corresponding generator. This action of $G$ on ${\cal{A}}\otimes F_{\infty}$ allows us to identify a weight on ${\cal{A}}\otimes F_{\infty}$ whose associated modular automorphism formally defines this "Hamiltonian". This is the {\it dominant weight} $\omega$ introduced in \cite{CT}. 

Once we reach this point we can can give meaning to the {\it the diff invariant ( with respect to modular time reparametrizations ) algebra of observables} as the centralizer $({\cal{A}}\otimes F_{\infty})_{\omega}$. This is equivalent to implement the standard {\it Hamiltonian gravity constraint}. The so defined algebra
$({\cal{A}}\otimes F_{\infty})_{\omega}$ is the type $II_{\infty}$ factor associated in \cite{Witten2} to the eternal black hole. Note that due to the KMS property we have defined a natural semifinite trace on this type $II_{\infty}$ factor determined by $\omega$.

Now we can define finite projections $P(s)$ in this type $II_{\infty}$ factor as well as one parameter families of type $II_1$ factors $P(s)({\cal{A}}\otimes F_{\infty})_{\omega}P(s)$. Moreover we can assign to each of these type $II_1$ factors the corresponding Murray von Neumann coupling. Thus we can model the {\it evaporation} of the black hole, defined at the time of formation by the type $III_1$ algebra ${\cal{A}}$, by the family of type $II_1$ factors associated with a one parameter family $P(s)$ of finite projections. In this model the coupling $d_B^{MvN}(s)$ is identified with the coupling of $P(s)({\cal{A}}\otimes F_{\infty})_{\omega}P(s)$.

To achieve our goal of constructing a unitary model of evaporation we need: i) to find the unitary operator $V_{s^{'}}$ formally satisfying (\ref{evapo}) for these $P_s$ and ii) to prove that the associated coupling $d_{B}^{MvN}(s)$ grows under its action. Fortunately there exist a natural candidate satisfying both conditions, namely {\it the generator of the dual automorphisms}.

In this sense {\it duality} identifies the two needed ingredients defining the evaporation process: the continuous dimension used to define the infinite dimensional generalization of $d_P(s)$ as well as the unitary evaporation operator. It is interesting to observe that this evaporation operator is not part of the diff invariant algebra.

Finally we could ask ourselves the most natural question namely: In what sense the so defined evaporation process is a {\it quantum} process ? The quantumness of evaporation is encoded in the {\it quantumness} of the Weyl-Heisenberg algebra  $F_{\infty}$ naturaly defined by the {\it double crossed product} i.e. by {\it duality}. The corresponding Planck constant defining this Weyl-Heisenberg algebra is crucial for the non trivial action of the evaporation operator. In more metaphoric terms: making this $\hbar$ equal zero prevents the evaporation process and makes the black hole eternal \footnote{This comment will become more clear in section 4.}. 

 \vspace{0.3 cm}

{\bf Remark} Unfortunately this note has turned out to be longer than it should have been. This is due to the author's ineffectiveness and the need to introduce technical aspects that are not very familiar. The reader can go directly to sections 3 and 4 and use section 2 and the appendix as supplementary material. In section 4 we insist on a model based on EPR pairs that is explained in the appendix. Also in the appendix we briefly discuss some aspects of the EPR/ER \cite{EPR} connection in the infinite dimensional context.

\section{Duality and crossed products}
\subsection{Covariance}
Let $M$ be a von Neumann algebra acting on a Hilbert space $H$ and let $G$ be a continuous locally compact abelian group acting on $M$ as
\begin{equation}
a\rightarrow \alpha_g(a)
\end{equation}
with $g\rightarrow \alpha_g$ a map from $G$ into $Aut(M)$, the full group of  automorphisms of $M$ \footnote{The map from $G$ into $M$ defined by $g\rightarrow \alpha_g(a)$ is assumed to be continuous in the sense that if $g_n$ converges to $g$ in the topology defining $G$ then $\alpha_{g_n}(a)$ converges, in the strong norm topology, to $\alpha_g(a)$ for any $a\in M$. }. As an example we can think of $G$ as time translations and $\alpha_g(a)$ the transformation of "observable" $a$ under the time translation defined by $g$ \footnote{In this case we can identify $G$ with ${\mathbb{R}}$.}. Thus we can replace $g$ by $t\in {\mathbb{R}}$ and $\alpha_g$ by $\alpha_t$. A priori we can distinguish two types of group actions, namely those for which the automorphism $\alpha_g$ is {\it inner} i.e. part of the algebra $M$, and those for which $\alpha_g$ is {\it outer}. 

Let us now define a {\it unitary} and continuous representation of $G$ on $H$ i.e. for the case of time translations, a continuous map $t\rightarrow U_t$ with $U_t$ a unitary operator acting on $H$. By a {\it covariant} implementation of the action of $G$ on $M$ we mean the existence of a unitary $U_t$ such that for any $a\in M$ we get
\begin{equation}\label{covariance}
\alpha_t(a)=U_taU_t^{*}
\end{equation}
In general not any group action of $G$ on $M$ defined by some $\alpha_g$ can be implemented covariantly. Moreover if $\alpha_t$ is outer any covariant implementation will imply that $U_t$ is not in $M$. In physics we will be interested in defining von Neumann operator algebras $M$ admitting:

 c-i) Covariant implementations of the continuous group of time translations and such that
 
 c-ii) The unitary $U_t$ is in $M$

Crossed products with respect to a group $G$ are, in essence, a general way to construct von Neumann algebras satisfying c-i) and c-ii) above.

In order to introduce the crossed product let us consider a von Neumann factor $M$ of type $III_1$ and let us define a weight $\phi$ on $M$\footnote{From now on we will assume that the weight is normal semi-infinite and faithful.}. Associated to this weight we can define the action of the group $G={\mathbb{R}}$ of "time translations" in terms of the modular automorphism $a\rightarrow \sigma^{\phi}_t(a)$ with $t\in G$. The GNS construction allows us to define the action of $M$ on the Hilbert space $H_{\phi}$ in terms of a representation $\pi_{\phi}$ of $M$ into the algebra $B(H_{\phi})$ of bounded operators acting on $H_{\phi}$. In order to define a covariant implementation of this modular action we need to find a unitary operator $U_t$ acting on $H_{\phi}$ such that (\ref{covariance}) \footnote{In what follows, and in order to simplify the notation, we will not make explicit the GNS representation $\pi_{\phi}$ replacing $\pi_{\phi}(a)$ by just $a$.}. In the case $M$ is of type $III$ we can use Tomita Takesaki modular operators $\Delta_{\phi}$ to represent $\sigma^{\phi}_t(a)=\Delta_{\phi}^{it} a \Delta_{\phi}^{-it}$. However, since in this case the modular automorphism is {\it outer}, the modular operator $\Delta$ is not part of the algebra $M$ and consequently is not satisfying condition  c-ii).  

The crossed product allows us to construct an algebra, associated to $M$, where the group $G$ of modular time translations can be implemented in a covariant way satisfying c-i) and c-ii) above \footnote{From now on by covariant implementation we will mean one satisfying c-i) and c-ii).}. The construction goes as follows:

 i) To extend the Hilbert space to $L^2(\mathbb{R},H_{\phi})$ defined by those functions $\psi(t)$ such that $\int dt ||\psi(t)||^2$ is finite and with $||\cdot||$ the norm defined in $H_{\phi}$ using the weight $\phi$, 
 
 ii) To define, on this space, a natural unitary representation of the group $G$ of modular time translations as $U_{t^{'}}\psi(t)=\psi(t-t^{'})$ and 
 
 iii) To define a representation $\pi_{\sigma}(a)$ of any $a\in M$ by a bounded operator acting on $L^2(\mathbb{R},H_{\phi})$ as
\begin{equation}\label{dress1}
(\pi_{\sigma}(a) \psi)(t)= \sigma_{-t} (a) \psi (t)
\end{equation}
where by $\sigma_{t}(a)$ we mean $\sigma^{\phi}_t(a)$. 

The crossed product algebra $M\rtimes_{\sigma}{\mathbb{R}}$ is now defined as {\it the algebra generated by $\pi_{\sigma}(a)$ and $U_t$ acting on $L^2(\mathbb{R},H_{\phi})$}.

In order to make contact with a more familiar physics notation let us denote $\pi_{\sigma}(a)$ as $\hat a$ and let us think $\hat a$ as a {\it gravitational dressing} of $a$ \cite{Witten0} \footnote{At this point the only reason to invoke {\it gravitational dressing} is that we are making covariant {\it modular time reparametrizations} in case we think $G$ as the topological group of modular time translations.} . Using (\ref{dress1}) we can formally represent $\hat a$ as
\begin{equation}
\hat a \psi(t)=\Delta_{\phi}^{-i\hat t} a \Delta_{\phi}^{i\hat t} \psi(t)
\end{equation}
provided we define 
\begin{equation}\label{time}
\Delta^{i\hat t} \psi(t)=\Delta^{it}\psi(t)
\end{equation}
with the formal operator $\hat t$ acting on $L^2(\mathbb{R})$ as a "position" operator i.e. $\hat t\psi(t)=t\psi(t)$. If we introduce a modular hamiltonian $\hat h_{\phi}$ by $\Delta_{\phi}^{it}=e^{it\hat h_{\phi}}$ we get the familiar expression \cite{Witten1} 
\begin{equation}
\hat a= e^{-i\hat t\hat h_{\phi}}ae^{i\hat t\hat h_{\phi}}
\end{equation}
Thus the unitary operator $U_t$ is defined as the generator of translations in $L^2(\mathbb{R})$. From these definitions follows the covariance relation namely 
\begin{equation}\label{U}
U_t\hat aU_t^{-1}=\widehat{a(t)}
\end{equation}
with $\widehat{a(t)}= e^{-i\hat t \hat h_{\phi}}(e^{-it\hat h_{\phi}}ae^{it\hat h_{\phi}})
e^{i\hat t \hat h_{\phi}}$. The unitary operator $U_t$ satisfying (\ref{U}) is given by $U_t=e^{i\hat p t}$ with $[\hat p,\hat t]=-i$. 

{\bf Remark 1} From a physics point of view we could think of the generator of $U_t$ as the {\it Hamiltonian} of the system described by the crossed product algebra $M\rtimes_{\sigma}{\mathbb{R}}$. Note that by contrast with the modular hamiltonian $h_{\phi}$, that is not part of the algebra $M$, now $U_t$ is part of the crossed product algebra. We will  discuss this important fact in section 2.4.

{\bf Remark 2} It is important to stress that the definition of the crossed product algebra $M\rtimes_{\sigma}{\mathbb{R}}$ {\it does not require the introduction of the operator $\hat t$}. Actually the generators of this algebra are $\hat a$ and $U_t$ with $\hat a$ defined as $\hat a\psi(t)=\sigma_{-t}^{\phi} \psi(t)$ with $t$ a c-number. The introduction of the conjugated transformation $V_s=: e^{is\hat t}$ requires {\it to add} the operator $\hat t$ conjugated to the operator $\hat p$ defining $U_t$. The meaning of an extended algebra where we add the operator $\hat t$ is related with the notion of {\it duality} that we will introduce in the next paragraph. Note also that the operators $\hat p$ and $\hat t$ are unbounded.

{\bf Remark 3} The definition of crossed product can be done for any locally compact abelian group G once the action on the algebra $M$ is defined by the corresponding automorphism $\alpha_g$. In this case the crossed product $M\rtimes_{\alpha}G$ is acting on $L^2(G,H)$ with $H$ the Hilbert space where $M$ is acting \footnote{And where we have equipped $G$ with a Haar measure.}.

{\bf Remark 4} Crossed products were initially defined by Murray and von Neumann in \cite{MvN1} using the {\it group measure construction}. In such a case we start with a measure space $(S,\mu)$ and we consider a numerable abelian group $G$ acting on the space $S$ preserving the measure $\mu$. The algebra $A$ of essentially bounded functions on $S$ i.e. $L^{\infty}(S,\mu)$ is acting on the Hilbert space $L^2(S,\mu)$ by left multiplication. The crossed product $A\rtimes G$, relative to the action of $G$ on $S$, is acting on the Hilbert space $L^2(G,H)$.

\subsection{The dual of modular transformations}
On the crossed product $M\rtimes_{\sigma}{\mathbb{R}}$ we can define a {\it dual action} that we can denote $\hat{\sigma}_s$ with $\hat{\sigma}_s$ defining a map $s\rightarrow \hat {\sigma}_s$ from the {\it dual group} $\hat G={\mathbb{\hat{R}}}$ into $Aut(M\rtimes_{\sigma}{\mathbb{R}})$. The action of $\hat{\sigma}_s$ is defined by
\begin{equation}\label{dualdef}
\hat{\sigma}_s(x)= \mu(s) x \mu(s)^{-1}
\end{equation}
for $x\in M\rtimes_{\sigma}{\mathbb{R}}$ with the {\it dual} action $\mu(s)$ defined as
\begin{equation}
\mu(s)=e^{is\hat t}=:V_s
\end{equation}
with $\hat t$ the "Heisenberg" dual to $\hat p$ for $\hat p$ defining the $U_t$ of $M\rtimes_{\sigma}{\mathbb{R}}$ \footnote{This is the operator $\hat t$ introduced in (\ref{time}).}. From (\ref{dualdef}) follows that $\hat \sigma_s(\hat a)=\hat a$ and that
\begin{equation}
\hat \sigma_s(U_t)= e^{ist} U_t
\end{equation}
and consequently the dual action defines an {\it automorphism} of $M\rtimes_{\sigma}{\mathbb{R}}$.
We will denote this dual action the {\it Heisenberg dual action} \footnote{Note that the conjugated operator $\hat t$ is not part of the algebra $M\rtimes_{\sigma}{\mathbb{R}}$ and therefore the dual action defines an {\it outer} automorphism.}. 

\subsection{Takesaki duality}
Once we have defined the {\it Heisenberg dual action} on $M\rtimes_{\sigma}{\mathbb{R}}$ we can define the {\it double crossed product}, namely
\begin{equation}\label{double}
(M\rtimes_{\sigma}{\mathbb{R}})\rtimes_{\hat \sigma}{\mathbb{\hat R}}
\end{equation}
acting now on $L^2({\mathbb{R}}\times {\mathbb{\hat R}},H_\phi)$. For the case of the group $G$ of modular time translations we identify elements in $G$ with $t\in{\mathbb{R}}$ and elements in the dual $\hat G$ with $s\in {\mathbb{\hat R}}$. Thus we can identify ${\mathbb{\hat R}}$ with ${\mathbb{ R}}$. The dual pairing $\langle s,t\rangle$ between $G$ and $\hat G$ being defined as $\langle s,t\rangle=e^{ist}$.

The algebra (\ref{double}) is generated by the {\it "dual dressing"} of the generators $\hat a$ and $U_t$ of $M\rtimes_{\sigma}{\mathbb{R}}$ and by $\mu(s)=:V_s$. The Takesaki duality
\begin{equation}
(M\rtimes_{\sigma}{\mathbb{R}})\rtimes_{\hat \sigma}{\mathbb{R}} = M\otimes F_{\infty}
\end{equation}
is telling us that this double crossed product is algebraically isomorphic to $M\otimes F_{\infty}$ where we can think $F_{\infty}$ as the algebra of bounded operators acting on $L^2(\mathbb{R})$. A basic result in quantum mechanics, due to Stone and von Neumann, \cite{SvN} is that $F_{\infty}$ is generated by the Heisenberg-Weyl algebra defined by $U_t$ and $V_s$ with the generators $\hat p$ and $\hat t$, satisfying $[\hat p,\hat t]=-i$. Consequently the Takesaki duality is telling us that the generators of the {\it double crossed product} are the elements of $M$ together with $U_t$ and $V_s$. 

\subsubsection{Coinvariance and the emergence of the observer}
Although the former conclusion on the set of generators of the double crossed product seems to be intuitive and natural the former duality is not as straightforward as it can looks at first sight. Indeed the algebra $M\otimes F_{\infty}$ is acting on the Hilbert space $H\otimes L^2(\mathbb{R})$ while the double crossed product is acting on $L^2(\mathbb{R}, L^2(\mathbb{R},H))= L^2({\mathbb{R}} \times {\mathbb{R}},H)$ \footnote{Note that we are identifying ${\mathbb{R}}$ and ${\mathbb{\hat R}}$.}. This implies the existence of a {\it "coinvariance"} map \footnote{See discussion in \cite{Takesaki} and in \cite{Witten2}.} 
\begin{equation}
T: L^2({\mathbb{R}})\otimes L^2({\mathbb{R}})\otimes H \rightarrow L^2({\mathbb{R}})\otimes H
\end{equation}
such that 
\begin{equation}
T a|\Psi\rangle = \tilde a T|\Psi\rangle
\end{equation}
with $a$ in the double crossed product acting on $L^2({\mathbb{R}} \times {\mathbb{R}},H)$ and $\tilde a$ in $M\otimes F_{\infty}$ acting on $L^2({\mathbb{R}})\otimes H$.

If we identify the type $I$ factor $F_{\infty}$ acting on $L^2(\mathbb{R})$ as formally defining a {\it quantum external observer} we discover that this observer naturally {\it emerges} once we impose: 

\vspace{0.3 cm}

{\it A covariant implementation of modular time automorphisms
 as well  as a covariant implementation of its dual action.}\footnote{Recall that by covariant we mean a covariant action satisfying c-i) and c-ii).}
 
 \vspace{0.3 cm}
 
 In the form of {\it slogan} we can put this result as
 
 \vspace{0.3 cm}
 
 {\it Covariant modular time translations $+$ Duality $=$ Quantum Observer}

\vspace{0.3 cm}

At this point it is important to stress the {\it quantum meaning} of duality. Note that the dual action $\mu(s)$ defines, relative to the $U_t$ used in the covariant implementation of modular time translations, a {\it quantum} Heisenberg-Weyl algebra with a non vanishing $\hbar$ that we have normalized to one.

\subsubsection{How modular time acts on the observer ?}
Once we reach this point the natural question should be to discover how the modular time automorphisms are acting on the emergent observer algebra. More precisely once we define the modular automorphism $\sigma_t^{\phi}$ acting on $M$ we want to define an extension $\tilde \sigma_t$ acting on $M\otimes F_{\infty}$. This action was defined in Takesaki's seminal paper \cite{Takesaki} as
\begin{equation}\label{actionobserver}
\tilde \sigma_t = \sigma_t^{\phi}
\otimes Ad (U_t)
\end{equation}
with the $U_t$ defined above \footnote{This is a non trivial result that involves the use of {\it coinvariance}. See theorem 4.6 and Lemma 6.6 in \cite{Takesaki}.}. In other words the modular time automorphisms are acting on the Hilbert space of the observer $L^2(\mathbb{R})$ as translations generated by $\hat p$. Hence what we can identify as {\it the observer momentum} i.e. $\hat p$ implements, on the observer Hilbert space, the modular time translations induced by the Takesaki duality. 

\subsubsection{Dominant weight}
Once we have identified how modular translations act on the combined system, that includes the {\it emergent} quantum observer, we can wonder for what weight $\omega$ on $M\otimes F_{\infty}$ the corresponding modular automorphism $\sigma_t^{\omega}$ will be precisely the $\tilde \sigma^{\omega}_t$ defined in (\ref{actionobserver}). The answer is clear from (\ref{actionobserver}) namely
\begin{equation}
\omega = \phi \otimes Tr(e^{\hat p},\cdot)
\end{equation}
The weight $\omega$ on $M\otimes F_{\infty}$ is a {\it dominant weight} in the sense of Connes-Takesaki \cite{CT} \footnote{For some physics discussion on the meaning of the dominant weight see \cite{Gomez2,Gomez3}.}. 

The {\it modular hamiltonian} $h_{\omega}$ formally defined
as $\tilde \Delta ^{it}=e^{it h_{\omega}}$ \footnote{With $\tilde \Delta$ the modular operator associated to $\tilde \sigma_t$.} is given by
\begin{equation}
h_{\omega}=h_{\phi}+\hat p
\end{equation}
and therefore the {\it centralizer} $(M\otimes F_{\infty})_{\omega}$ is defined by the elements in $M\otimes F_{\infty}$ {\it invariant} under the action of the modular time translations defined in (\ref{actionobserver}). More precisely
\begin{equation}\label{conmutant1}
(M\otimes F_{\infty})_{\omega}= M\otimes F_{\infty} \cap \{U_t\otimes \sigma_t^{\phi} \}^{'}
\end{equation}

Note that $(M\otimes F_{\infty})_{\omega}$ implements the GR hamiltonian constraint for $h_{\omega}$.
In metaphoric language we can again summarize the former discussion in the form of the slogan:

\vspace{0.3 cm}

{\it Invariance under double duality $\Rightarrow$ GR diff invariance}

\vspace{0.3 cm}

where by GR diff invariance we mean those observables commuting with $U_t\otimes \sigma_t^{\phi}$.

\subsection{Example: the eternal black hole}
As an example let us consider the Penrose diagram in Figure 4 and identify $M$ as the type $III_1$ factor associated with the $R$ wedge. Denote this algebra $A_R$ \footnote{This is the algebra we denoted ${\cal{A}}$ in the introduction.}. Once a weight $\phi$ on $A_R$ is introduced we can use {\it covariance} to define the crossed product algebra algebra $A_R\rtimes_{\sigma_t^{\phi}} {\mathbb{R}}$. Let us denote this algebra $\cal{A}_{R}$. 
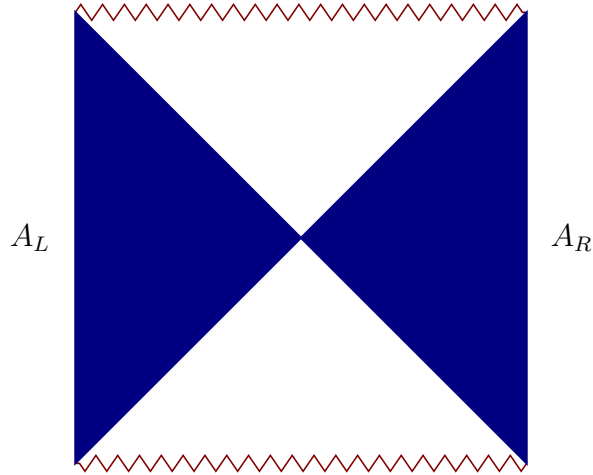
\begin{figure}[htbp]  
 \begin{center}
   \begin{tikzpicture}[scale=3]   
   \def\R{0.08} 
   \def\Nlines{3} 
   \pgfmathsetmacro\ta{1/sin(90*1/(\Nlines+1))} 
   \pgfmathsetmacro\tb{sin(90*2/(\Nlines+1))} 
   \pgfmathsetmacro\tc{1/sin(90*2/(\Nlines+1))} 
   \pgfmathsetmacro\td{sin(90*1/(\Nlines+1))} 
   \coordinate (-O) at (-1, 0); 
   \coordinate (-S) at (-1,-1); 
   \coordinate (-N) at (-1, 1); 
   \coordinate (-W) at (-1, 0); 
   \coordinate (-E) at ( 0, 0); 
   \coordinate (c1) at ( 0, 1); 
   \coordinate (c2) at ( 00, -1); 
   \coordinate (O) at ( 1, 0); 
   \coordinate (S) at ( 1,-1); 
   \coordinate (N) at ( 1, 1); 
   \coordinate (E) at ( 2, 0); 
   \coordinate (W) at ( 0, 0); 
   \coordinate (B) at ( 0,-1); 
   \coordinate (T) at ( 0, 1); 

   \begin{scope}
   
   \clip[decorate,decoration={zigzag,amplitude=2,segment length=6.17}]
   (S) -- (-S) --++ (-1.1,-0.1) |-++ (4.2,2.2) |- cycle;
   \clip[decorate,decoration={zigzag,amplitude=2,segment length=6.17}]
   (-N) -- (N) --++ (1.1,0.1) |-++ (-4.2,-2.2) |- cycle;
 
   \end{scope}
   
   \draw[singularity] (-N) -- node[above] {} (N);
   \draw[singularity] (S) -- node[below] {} (-S);
   \path (S) -- (W) node[myblue,pos=0.6,below=-2.5,rotate=-45,scale=0.85]
   {};
   \path (W) -- (N) node[myblue,pos=0.32,above=-2.5,rotate=45,scale=0.85]
   {};
   \draw[thick,myblue] (-N) -- (-E) -- (-S) -- cycle;
   \draw[thick,myblue] (N)  -- (S) -- (W) -- cycle;
   \fill[thick,myblue,opacity=0.3] (N)  -- (S) -- (W) -- cycle;
   \fill[thick,myblue,opacity=0.3] (-N) -- (-E) -- (-S) -- cycle;
   \draw (-1.2,0.0) circle(0pt) node[] {$A_L$};
   \draw (1.2,0.0) circle(0pt) node[] {$A_R$};
 \end{tikzpicture} 
\end{center}
 \caption{Penrose Diagram of the double-sided AdS Schwarzschild Black Hole.}
 \label{fig:4} 
\end{figure}   

The algebra associated to the $L$ wedge is given by the corresponding commutant
\begin{equation}
{\cal{A}}_{L}= (\cal{A}_{R})^{'}
\end{equation}
From (\ref{conmutant1}) it follows that ${\cal{A}}_{L}$ is generated by $A_R^{'}\otimes {\mathbb{C}}$ and $h_{\phi}+\hat p$. This means that $(h_{\phi}+\hat p)\in {\cal{A}}_{L}$.

Since $A_R^{'}\otimes {\mathbb{C}}$ is in ${\cal{A}}_{L}$ it superficially looks that we can define the $L$ algebra $A_R^{'}$ without any {\it gravitational dressing}. This is a mirage. Indeed once we implement modular time translations on the $L$ algebra $A_R^{'}$ in a {\it covariant} way we need to use a transformation $U^{'}_t$ ( the commutant analog of $U_t$ in $\cal{A}_{R})$ ) with generator $h_{\phi}+\hat p$ that automatically dress the elements in 
$A_R^{'}$\footnote{For a more technical discussion see Corollary 5.13 in \cite{Takesaki}.}. 

\subsubsection{Hamiltonians and energetics}
For the Penrose diagram of the eternal black hole in Figure 4 imagine yourself as the asymptotic $R$ observer. If you intend to do quantum field theory in this background you will use, as algebra of observables, an algebra associated to the wedge $R$ in the figure i.e. an algebra of local observables with support in $R$. This is the algebra we denoted $A_R$ in the previous section \footnote{And simply as ${\cal{A}}$ in the introduction.}. Normally you will like to define the algebra $A_R$ as an algebra of operators acting on some Hilbert space $H$ of quantum states and to identify the different experimental outputs of your local experiments with different {\it states} ( or more general {\it weights}) i.e. with different linear forms on $A_R$. A more sophisticated assumption, reflecting the standard quantum mechanics intuition about the notion of quantum fluctuation around a given classical background, will be to associate with a given {\it state} $\phi$ a "vacuum" state $|0,\phi\rangle$ and to identify the elements in $A_R$ with the different states representing quantum fluctuations on the vacuum $|0,\phi\rangle$. Now, thinking perturbatively, you can identify $H$ as the space linearly generated by states of the type $a|0,\phi\rangle$ with $a\in A_R$. After completion this defines the GNS Hilbert space $H_{\phi}$. Now, and once we have defined the action of $A_R$ on $H_{\phi}$, you can define the corresponding commutant $A_R^{'}$ that we have denoted $A_L$ in the previous section. If we keep going with the Penrose diagram of Figure 1 we can think $A_L$ as the algebra of local operators describing quantum fluctuations, relative to the background defined by $\phi$, but localized in the $L$ wedge of the diagram. Now let us make the big assumption that $A_R$ is a von Neumann factor and that $A_R \cap A_L={\mathbb{C}}1$ \cite{Witten2}. Next and based on Araki's basic result  on the nature of von Neumann algebras of local observables in bounded regions of space-time \cite{Araki}, let us assume that $A_R$ is a type $III_1$ factor. Reaching this point we pose the following general question: 

\vspace{0.3 cm}

{\it What will qualify for this asymptotic observer as a Hamiltonian, let us say $H_R$ ?}

\vspace{0.3 cm}

Our former discussion on {\it covariance} naturally leads to an answer based on the following steps: i) Define the modular time automorphism $\sigma_t^{\phi}$ acting on $A_R$, ii) Define the {\it covariant} representation of modular time translations on the crossed product $A_R\rtimes_{\sigma_t}{\mathbb{R}}$ acting on $L^2({\mathbb{R}},H_{\phi})$ and iii) Identify the desired Hamiltonian $H_R$ with the generator of modular time translations $U_t$ on $L^2({\mathbb{R}},H_{\phi})$ making the modular action covariant i.e. with the operator $\hat p$ defined in previous sections. 

Thus the {\it Hamiltonian $H_R$} becomes a well defined element in $A_R\rtimes_{\sigma_t}{\mathbb{R}}$. In order to appreciate the physics meaning of this construction you should be aware that no element in $A_R$ defines the effect of time translations on $A_R$\footnote{Obviously this will not be the case if we work with $A_R$ a type $I$ factor and with $H$ a finite dimensional Hilbert space.}. What about the $L$ observer ? If the $L$ observer follows the same logic to define $H_L$ she will conclude that the candidate for $H_L$ is $\hat p +h_{\phi}$ that, as discussed, is now an element of ${\cal{A}}_{L}$.

The crucial difference between $H_R$ and $H_L$ is that while $H_R$ is totally independent on the {\it background} defined by $\phi$ this is not the case for $H_L$. Using these definitions of the Hamiltonians $H_L$ and $H_R$ we easily observe that
\begin{equation}
h_{\phi}=H_L-H_R
\end{equation}
with $H_L$ and $H_R$ {\it well defined operators} in ${\cal{A}}_{L}$ and ${\cal{A}}_R$ respectively.

\subsubsection{Energetics}
Once we have defined $H_R$ and $H_L$ we should expect the asymptotic observer i.e. the $R$ observer, could be interested in defining the average {\it energy} of the state representing what she is observing. As discussed this state will be represented by some weight $\Phi$ on ${\cal{A}}_R$ thus the corresponding average energy of this state will be
\begin{equation}
\Phi(H_R)
\end{equation}
As discussed in previous section we can define , using {\it duality} and the double crossed product, the dominant weight $\omega$ on $A_R\otimes F_{\infty}$ and to identify ${\cal{A}}_{R}$ with the corresponding centralizer i.e.
\begin{equation}
(A_R\otimes F_{\infty})_{\omega}={\cal{A}}_{R}
\end{equation}
Using now the KMS property for $\omega$ we can define a trace weight on ${\cal{A}}_R$, let us say $\tau_{\omega}$, by simply reducing $\omega$ to the centralizer $(A_R\otimes F_{\infty})_{\omega}$.  

In terms of $\tau_{\omega}$ we can define, associated to any state ( weight ) $\Phi$, the {\it affiliated density matrix} $\rho_{\Phi}$ as a self adjoint operator affiliated to ${\cal{A}}_R$ defined by 
\begin{equation}
\Phi(a) =\tau_{\omega}(\rho_{\Phi}a)
\end{equation}
for any $a\in {\cal{A}}_R$. Thus the desired definition of $R$ energy for an arbitrary state $\Phi$ is given by
\begin{equation}
\Phi(H_R)=\tau_{\omega}(\rho_{\Phi}H_R)
\end{equation}
We will denote this quantity as ${\cal{E}}^{R}(\Phi)$.

\subsection{Dilatations and Duality}
In the previous section we have shown the isomorphism
\begin{equation}
(A_R\otimes F_{\infty})_{\omega}= A_R\rtimes_{\sigma_t^{\phi}} \mathbb{R}=:{\cal{A}}_R
\end{equation}
with $\omega$ the dominant weight defined by {\it duality} and double crossed product. We stressed that on the centralizer $(A_R\otimes F_{\infty})_{\omega}$ we can define, using the KMS property, a trace $\tau_{\omega}$. Now we want to see how this trace transforms under the {\it dual action} $\hat\sigma_s$ defined in section 2.2.  
More precisely we want to discover how the trace $\tau_{\omega}(a)$ for $a\in {\cal{A}}_R$ is related with the trace of $\hat \sigma_s(a)$.

 Let us focus on the Hamiltonian $H_R$ that, as discussed in the previous section, is in ${\cal{A}}_R$. The "energy" for the asymptotic $R$ observer of the "ground" state associated to the dominant weight $\omega$ was defined as
\begin{equation}
{\cal{E}}^{R}(\tau_{\omega}) = \tau_{\omega}(H_R)
\end{equation}
Using the definition in section 2.2 of the dual action $\hat\sigma_s$ we easily get \footnote{See theorem 1.3 in \cite{CT}.}
\begin{equation}
\tau_{\omega}(\hat \sigma_s(H_R))=e^{-s}\tau_{\omega}(H_R)
\end{equation}
Meaning that the "energy" ${\cal{E}}^{R}(\tau_{\omega})$ as measured by the asymptotic $R$ observer {\it scales} under dual modular transformations. It is important to stress that this "dilatation" of the "energy" under dual transformations follows from the Heisenberg-Weyl relation defining the algebra generated by $U_t,V_s$ used to make covariant the modular action $\sigma_t$ and its dual $\hat\sigma_s$ respectively.

The precise mathematical meaning of the former result is that the trace weight $\tau_{\omega}$ is {\it relatively invariant} \cite{Takesaki} under the dual transformations defined by $\hat\sigma_s$. This allows us to define the crossed product
\begin{equation}
{\cal{A}}_R\rtimes_{\hat \sigma_s}{\mathbb{R}}
\end{equation}
and to identify the dominant weight $\omega$ on $A_R\otimes F_{\infty}$ with the {\it dual} ( in Takesaki terminology ) of the trace weight $\tau_{\omega}$.

In words we can summarize the former discussion saying:

\vspace{0.3 cm}

{\it The dual to modular time translations  induces dilatations of the asymptotic energy.}

\vspace{0.3 cm}

For any other weight $\Phi$ on ${\cal{A}}_R$ we define the "energy" as ${\cal{E}}^{R}(\Phi)=\tau_{\omega}(\rho_{\Phi} H_R)$. Note that these "energies" depend on the state $\phi$ on $A_R$ used to define the GNS Hilbert space as {\it the space of quantum fluctuations on the background defined by $\phi$}.
\subsubsection{Action of duality on weights}
Recall that given a generic weight $\Phi$ on the centarlizer $(A_R\otimes F_{\infty})_{\omega}$ we define the affiliated {\it density matrix} $\rho_{\Phi}$ by $\Phi(x)=\tau_{\omega}(\rho_{\Phi}x)$. In addition we know that under dual modular transformations $x\rightarrow \hat\sigma_s(x)$ for $x\in (A_R\otimes F_{\infty})_{\omega}$ we have 
\begin{equation}
\tau_{\omega}(\hat \sigma_s(x)) = \lambda(s) \tau_{\omega}(x)
\end{equation}
with $\lambda(s)=e^{-s}$. Let us now define the action of the dual automorphisms on the space of weights i.e. let us define $T_s:\Phi\rightarrow \Phi^s$ such that
\begin{equation}
\Phi^s(x) = \tau_{\omega}(\hat \sigma_s(\rho_{\Phi}x))
\end{equation}
It is an easy exercise to show that
\begin{equation}\label{transformation}
\Phi^s(x)=\lambda(s)\Phi(x)
\end{equation}
again for any $x\in (A_R\otimes F_{\infty})_{\omega}$. This leads to the scale transformation of the "energy" under dual transformations. Indeed for any weight $\Phi$ the energy ${\cal{E}}^{R}(\Phi)= \Phi(H_R)$ transforms as
\begin{equation}
{\cal{E}}^{R}(\Phi)\rightarrow {\cal{E}}^{R}(\Phi^s) =\lambda(s){\cal{E}}^{R}(\Phi)
\end{equation}

\subsection{Some remarks on the used notion of "energy" and its relation to entropy}
We can interpret ${\cal{E}}^{R}(\tau_{\omega})$ as the $R$ energy of the {\it background} defined by the state $\phi$ used to define $\omega$. This is not a {\it physical} energy, since $H_R$ is not positive, and that is the reason we used quotations. The same for ${\cal{E}}^{R}(\Phi)$ for $\Phi$ a weight on ${\cal{A}}_R$ different from $\tau_{\omega}$. We can think of ${\cal{E}}^{R}(\Phi)$ as representing the $R$ "energy" of some excited state represented by the weight $\Phi$. 

Let us now try to unveil the meaning of ${\cal{E}}^{R}(\tau_{\omega})$. Note that the weight $\tau_{\omega}$ is uniquely defined once we have fixed the {\it background} $\phi$. 

Recall from section 2.3.3 that the weight $\omega$ on $A_R\otimes F_{\infty}$ is defined as $\omega=\phi\otimes \bar \omega$ with $\bar\omega$ the weight on $F_{\infty}$ defined by
\begin{equation}
\bar \omega(x)=Tr(e^{H_R} x)
\end{equation}
for $x\in F_{\infty}$. We can map $x$ into $1\otimes x$ in $A_R\otimes F_{\infty}$, thus $\omega(x)=\phi(1). Tr(e^{H_R} x)$ and consequently
\begin{equation}
{\cal{E}}^{R}(\tau_{\omega})=\omega(H_R) = Tr(e^{H_R} H_R)
\end{equation}
with $Tr$ the standard trace on $F_{\infty}$. Note that this quantity is {\it divergent} since the spectrum of $H_R=\hat p$ as an operator on $L^2(\mathbb{R})$ is the whole ${\mathbb{R}}$. What about ${\cal{E}}^{R}(\Phi)$ ? 

This quantity is defined as
\begin{equation}
{\cal{E}}^{R}(\Phi)= Tr(e^{H_R}\rho_{\Phi} H_R)=\Phi(H_R)
\end{equation}
and the transformation under dual automorphisms $\hat\sigma_s$ is given by ${\cal{E}}^{R}(\Phi)\rightarrow {\cal{E}}^{R}(\Phi^s)=\lambda(s){\cal{E}}^{R}(\Phi)$.

Let us now define the {\it entropy} ${\cal{S}}^{R}(\Phi)$ as \cite{Longo}

\begin{equation}
{\cal{S}}^{R}(\Phi)= -\tau_{\omega}(\rho_{\Phi}\log(\rho_{\Phi})) = -\Phi(\log\rho_{\Phi})
\end{equation}

We have denoted this entropy ${\cal{S}}^{R}$ to indicate that the density matrix $\rho_{\Phi}$ is affiliated to the centralizer $(A_R\otimes F_{\infty})_{\omega}$.

How this {\it entropy} transforms under dual transformations $\hat \sigma_s$ ? The transformation could be defined in two ways \cite{Witten2}, namely as 
\begin{equation}\label{case1}
{\cal{S}}^R(\Phi)\rightarrow {\cal{S}}^R(\Phi^s)
\end{equation}
and as
\begin{equation}
{\cal{S}}^R(\Phi)=-\tau_{\omega}(\rho_{\Phi}\log \rho_{\Phi})\rightarrow -\tau_{\omega}(\hat \sigma_s(\rho_{\Phi}\log \rho_{\Phi}))
\end{equation}
In the second case we get formally (i.e. assuming $\log\rho_{\Phi}$ is an affiliated operator) that ${\cal{S}}(\Phi)$ scales with $\lambda(s)$ under modular dual transformations. In the case (\ref{case1}) we get instead
\begin{equation}
{\cal{S}}(\Phi)\rightarrow \lambda(s) ({\cal{S}}(\Phi)+\log(\lambda(s)))
\end{equation}
\subsubsection{An  "observer" entropy}
For the weight $\bar \omega$ on $F_{\infty}$ used to define the dominant weight $\omega$ as $\phi\otimes \bar\omega$ we can define an affiliated density matrix by
\begin{equation}
\bar\omega(x)=Tr(\tilde \rho_{\bar\omega} x)
\end{equation}
with $x\in F_{\infty}$ and $Tr$ the standard trace on $F_{\infty}$. In this case we get $\tilde \rho_{\bar\omega}=e^{H_R}$. This allows us to define an {\it observer entropy} \footnote{In case we interpret $F_{\infty}$ as the type $I_{\infty}$ factor representing the observer.} as 
\begin{equation}
\tilde {\cal{S}}^{R}(\tilde\rho_{\bar{\omega}})=- Tr(\tilde\rho_{\bar\omega}\log \tilde \rho_{\bar \omega})
\end{equation}

Obviously $\tilde S^{R}(\tilde\rho_{\bar \omega})=-\bar \omega(H_R)$. Thus we get
\begin{equation}\label{energyentropy}
{\cal{E}}^R(\tau_{\omega})=-\tilde {S}^R(\tilde\rho_{\bar \omega})
\end{equation}

\subsubsection{The $L$ version}
In the same way we have defined ${\cal{E}}^R(\Phi)$ and ${\cal{S}}^R(\Phi)$ for an arbitrary weight $\Phi$ on $(A_R\otimes F_{\infty})_{\omega}$ we could define the analog $L$ quantities if we use the commutant of $(A_R\otimes F_{\infty})_{\omega}$ that we can identify as ${\cal{A}}_L$. Recall that this commutant is generated by $A_L\otimes {\mathbb{C}}$ and $U_t\otimes \Delta_{\phi}^{it}$ leading to define $H_L=h_{\phi}+H_R$ \footnote{We can use the convention of replacing $H_R$ by $-H_R$ avoiding the minus sign in (\ref{energyentropy}). In this convention we get the most familiar representation $h_{\phi}=H_R-H_L$.}. 

Thus ${\cal{E}}^L(\Phi)= \Phi(H_L)$ and
\begin{equation}
{\cal{S}}^L(\Phi) = \tau_{\omega}(\rho^L_{\Phi}\log \rho^L_{\Phi})
\end{equation}
with $\rho_{\Phi}^L=J\rho_{\Phi}J$ an affiliated operator of ${\cal{A}}_L$ and $J$ the corresponding antilinear Tomita Takesaki operator.

It is easy to see that for ${\cal{A}}_R$ and ${\cal{A}}_L$ defined above we get
\begin{equation}
{\cal{S}}^L(\Phi)={\cal{S}}^R(\Phi)
\end{equation}
for any weight $\Phi$. 

Unfortunately for the case of the type $II_{\infty}$ factors ${\cal{A}}_R$ and ${\cal{A}}_{L}$ the quantities ${\cal{S}}^{R,L}$ as well as ${\cal{E}}^{L,R}$ are infinite. 
\footnote{In \cite{Witten2} a {\it time shift} {\it operator}, let us call it $\Delta$, was introduced. In our present notation this time shift operator is $\hat t$, the Heisenberg conjugated of $H_R$. Note that this operator can be identified with the generator of the dual automorphism $\hat\sigma_s$ acting on $A_R\rtimes_{\sigma_t}{\mathbb{R}}$ for $\sigma_t$ the standard modular automorphism acting on $A_R$. 

Thus we can conclude that :

{\it time shift operator $=$ generator of the dual modular automorphisms.}}

\subsection{Why the external observer is quantum?}
In the previous section we have shown how {\it Takesaki duality} leads to the {\it emergent} observer algebra $F_{\infty}$ generated by the Weyl-Heisenberg algebra defined by $U_t,V_s$. The way {\it duality} is related to {\it quantum mechanics} is a basic consequence of Stone von Neumann theorem. In a nutshell let us consider the group $G={\mathbb{R}}$ and its dual $\hat G$. A pair of covariant representattions $U_t$ and $V_s$ of $G$ and $\hat G$ define the algebra
\begin{equation}
U_tV_sU_t^{-1}=\langle s,t\rangle V_s
\end{equation}
that is thanks to Stone von Neumann theorem equivalent to a {\it multiple} of the standard Schrodinger representation on $L^2(\mathbb{R})$ of the Heisenberg commutation relation $[\hat p,\hat t]=-i\hbar$. Morally speaking we introduce a non vanishing $\hbar$ once we implement {\it covariantly} both $G$ and its dual $\hat G$ ( for $G$ the group of modular time translations ).

The {\it scale} $\hbar$ can be made explicit explicitly defining $\langle s,t\rangle=: e^{ist\hbar}$. In this case we can visualize the scaling factor of the formal energy defined above as $e^{-s\hbar}$. 

It is important to stress the conceptual difference between the {\it quantumness} of the algebra $A_R$ that is a non abelian algebra of operators acting on a Hilbert space of states and the {\it quantumness} of the emergent $F_{\infty}$ factor i.e. the nonvanishing value of the $\hbar$ defining the Weyl-Heisenberg algebra. This {\it observer quantumness} is a direct consequence of {\it duality}.

\section{Duality and BH evaporation: towards an algebraic solution to the information paradox}

\subsection{Heuristic approach}
In a nutshell Page's approach to the information paradox, as already briefly discussed in the introduction, follows from assuming that the process of evaporation can be described in terms of a one parameter family of bipartite decompositions of a Hilbert space $H$ of {\it finite dimension}, let us say $N$, where each factorization $H=H_B(s)\otimes H_{R}(s)$ accounts for the Hilbert space of states of the black hole at that moment of evaporation $H_B(s)$ as well as the Hilbert space of the emitted radiation $H_R(s)$. In this setup we can describe the evaporation process in terms of 
\begin{equation}\label{dimensions}
d_P(s)=\frac{d(H_R(s))}{d(H_B(s))}
\end{equation}
with $d(H_B(s))$ and $d(H_R(s))$ representing the dimensions of the corresponding finite dimensional  Hilbert spaces. In this finite dimensional context the full evaporation process is characterized by a function $d_P(s)$ that starts with value $O(\frac{1}{N})$, corresponding to the moment of black hole formation, and ends with a value $O(N)$ at the final moment of evaporation ( see Figure 1). 

Denoting as $\Psi(s)$ the pure quantum state of the full system along the process, we can define the entanglement entropy $S(s,\Psi(s))$ as the von Neumann entropy for the corresponding bipartite decomposition. If at each evaporation time $s$ the state $\Psi(s)$ maximizes the entanglement entropy we will find a Page curve for the entropy defined by 
\begin{equation}\label{Pageentropy}
S_P(s)=\min\{\log d(H_B(s)), \log d(H_R(s))\}
\end{equation}
 Thus $S_P(s)$ will start at the value $0$, corresponding to the initial time with $d(H_R(s))=1$ and will end also with the value $0$ corresponding to the end of evaporation with $d(H_B(s))=1$. The maximum of this entropy will be reached at the Page time $s_P$ where $d(H_B(s))=d(H_R(s))=\sqrt{N}$ ( see Figure 2). If we use a {\it random} state $\Psi(s)$ we can define $S_P(s)$ as the {\it average} of $S(s,\Psi(s))$. The formal Page curve defined in (\ref{Pageentropy}) for $\Psi(s)$ maximally entangled at all values of $s$ defines formally the enveloping Page curve.

In addition to $S_P(s)$ we can define the {\it information functions} $I^{B,R}(s)$ by \cite{Page}
\begin{equation}\label{information}
I^{B,R}(s,\Psi(s))=\log d(H_{B,R}(s))-S(s,\Psi(s))
\end{equation}
We can consider this information function for an arbitrary random state $\Psi(s)$ and to evaluate the average value. This was done in \cite{Page1}. For $\Psi(s)$ the maximally entangled state, we get that for $s<s_P$
\begin{equation}
I^R(s)=0
\end{equation}
while for $s>s_P$ we get
\begin{equation}
I^R(s)= \log\frac{dH_R(s)}{dH_B(s)}= \log d_P(s)
\end{equation}
for $d_P(s)$ defined in (\ref{dimensions}). This increase of the information in the radiation, after Page's time $s_P$, reflects how, in this setup, the state is purified. In \cite{Page1} it was shown that this increase of information takes place for the {\it average} information for a random state.

Note that in this formal solution of the information paradox it is crucial to assume that the full Hilbert space $H$ has a finite dimension $N$. Indeed the information paradox is solved provided $d_P(s)$ grows monotonically from an initial value $O(\frac{1}{N})$ to a final value $O(N)$. Under this condition the value of $S_P(s)$ follows, for a typical state, a Page curve of the type depicted in (Figure 2). Obviously the information paradox will appear if $d_P(s)$ stops growing after reaching the value $1$ (see (Figure 5 and 6)).
\begin{figure}
    \begin{center}
        \begin{tikzpicture}[scale=2.95]
        \draw[dashed]    (-0.01,1.1)--(2.0,1.1);
        \draw[-stealth,thick,line width=1.75](0,-.02)--(2.35,-.02)node[anchor=north]{$s$};
        \draw[-stealth,thick,line width=1.75](0,-.02)--(0,1.8)node[anchor=east]{$S_p(s)$};
        \filldraw(0,0) circle(.25pt);
        \draw[thick]  (-0.05,1.15)circle(0pt) node[anchor=east ] {$1$};
        \draw[mydarkred,line width=1.5] (0,0)..controls+(.22,.45) and +(-.1,.01)..(1.8,1.005);
        \draw[mydarkred,line width=1.5]  (1.78,1.003)--(1.9,1.02) node[anchor=north ] {};    
    \end{tikzpicture}
    \end{center}
    \caption{An example leading to the information paradox.}\label{fig:5} 
\end{figure}
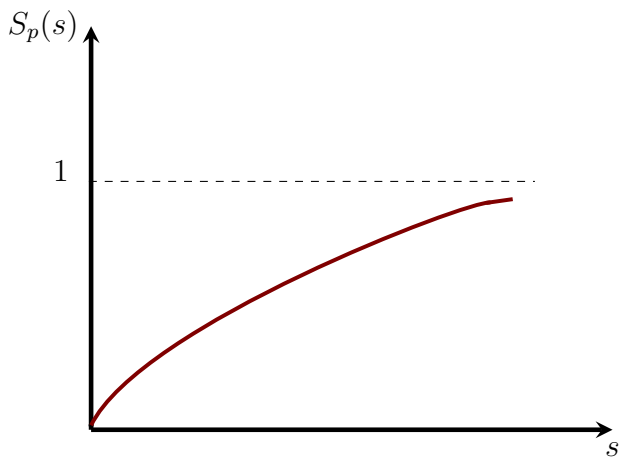
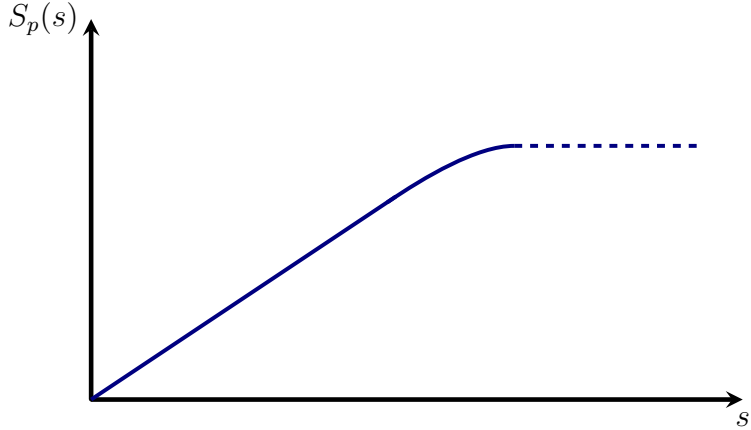
\begin{figure}
    \begin{center}
        \begin{tikzpicture}[scale=2.25]
        \draw[-stealth,thick,line width=1.75](0,0)--(3.85,0)node[anchor=north]{$s$};
        \draw[-stealth,thick,line width=1.75](0,0)--(0,2.25)node[anchor=east]{$S_p(s)$};
        \filldraw(0,0) circle(.25pt);
        \draw[myblue,line width=1.5]  (0,0)--(1.8,1.2) node[anchor=north ] {};
        \draw[myblue,line width=1.5] (1.75,1.166)..controls+(.22,.15) and +(-.22,.0)..(2.5,1.5);
        \draw[myblue,line width=1.5,dashed]  (2.5,1.5)--(3.6,1.5) node[anchor=north ] {};    
    \end{tikzpicture}
    \end{center}
    \caption{Cartoon of the entropy for the $d_P(s)$ of Figure 5}\label{fig:6} 
\end{figure}

Let us now pose the question:

\vspace{0.3 cm}

{\it Can we generalize this sort of solution of the information paradox when we work with an infinite dimensional Hilbert space $H$?}

\vspace{0.3 cm}

The answer as described in \cite{Gomez1} is yes. Before entering into technicalities let us recall the basic elements leading to this conclusion.

\subsection{Basic algebraic steps}
{\bf Step 1} Instead of describing the evaporation process by the one parameter family of bipartite decompositions of the full finite dimensional Hilbert space $H$ we will describe the evaporation process as:

\vspace{0.3 cm}

{\it A one parameter family of couples of {\bf finite} factors $(M_B(s),M_R(s))$, acting on an infinite dimensional Hilbert space and such that
\begin{equation}\label{decomposition}
M_R(s)=M_B(s)^{'}
\end{equation}

and consequently $M_B(s)\cap M_R(s)={\mathbb{C}}1$} 
\vspace{0.3 cm}

We will interpret $M_B(s)$ and $M_R(s)$ as the finite factors describing at each evaporation time $s$ the black hole and the radiation entanglement wedges respectively \footnote{By $M_B(s)^{'}$ we mean the commutant of $M_B(s)$.}. Note that since we are describing each evaporation time using a {\it finite factor} and its commutant we are already including the information about the {\it infinite dimensional} Hilbert space on which these factors are acting. Thus, our first step is to replace Page's family of bipartite decompositions of a finite dimensional Hilbert space, by a family of couples of finite factors satisfying (\ref{decomposition}). 

\vspace{0.3 cm}

{\bf Step 2} The next step will consist in defining the {\it infinite dimensional} analog of (\ref{dimensions}). If $M_B(s)$ and $M_R(s)$ are finite factors acting on an infinite dimensional Hilbert space they should be type $II_1$ factors. This means that we count with two well defined generalized dimensions: $d_B$ and $d_R$ from the space of projections ${\cal{P}}_B$ and ${\cal{P}}_R$ into the interval $[0,1]$ \footnote{After appropriated normalization.}. Here ${\cal{P}}_B$ and ${\cal{P}}_R$ denote the space of projections in $M_B(s)$ and $M_R(s)$ respectively. Recall that for any finite projection $p\in M_B(s)$ the quantity $d_B(p)$ defines the generalized dimension of the subspace of the Hilbert space $H$ defined by $p$ i.e. $pH$. 

With these basic ingredients we can define the {\it continuous} analog of (\ref{dimensions}) as
\begin{equation}\label{basicdefinition}
d_B^{MvN}(s)=:\frac{d_B(M_R(s)\psi(s))}{d_R(M_B(s)\psi(s)}
\end{equation}
or equivalently $d_R^{MvN}(s)=\frac{1}{d_B^{MvN}(s)}$.

Let us see the physics meaning of this quantity. At each evaporation time $s$ we define for a generic quantum vector state $\psi(s)$ \footnote{Where by that we mean a vector state in the Hilbert space on which the couple $(M_B(s),M_R(s))$ are acting.} the subspace $M_R(s)\psi$ i.e. all states in the Hilbert space that can be obtained acting with elements of $M_R(s)$ i.e. with the algebra of the radiation wedge factor, on the state $\psi(s)$. Since we are assuming (\ref{decomposition}) the projector associated with the subspace $M_R(s)\psi(s)$ is an element of $M_B(s)$ and therefore we can evaluate the generalized dimension of this space using the continuous dimension $d_B$. Thus the numerator of (\ref{basicdefinition}) defines the continuous analog of the dimension of the radiation Hilbert space $H_R(s)$ in the finite bipartite decomposition. However in this continuous case this quantity depends on the particular quantum state $\psi(s)$. The same can be said about the denominator of (\ref{basicdefinition}) that represents the continuous analog of the dimension of the Hilbert space $H_B(s)$ describing, in the bipartite picture, the black hole Hilbert space at time $s$. 

The quantity $d_B^{MvN}(s)$ defines  the continuous version of Page's function $d_P(s)$ defined in (\ref{dimensions}). This quantity has a very interesting algebraic meaning, namely is the Murray von Neumann {\it coupling} for the type $II_1$ factor $M_B(s)$. 

Obviously we could also define the {\it dual} version $d_R^{MvN}(s)=\frac{1}{d_B^{MvN}(s)}$ just replacing the roles of $M_B$ and $M_R$. 

\vspace{0.3 cm}

{\bf Step 3} A basic result of Murray and von Neumann is that the coupling $d_B^{MvN}(s)$ defined in (\ref{basicdefinition}) is {\it independent of the state $\psi(s)$}. In other words, the value of $d_B^{MvN}(s)$ only depends on the family of finite factors $(M_B(s),M_R(s))$ but not on the particular state $\psi(s)$ used in (\ref{basicdefinition}). Obviously the same is true for its inverse $d_R^{MvN}(s)$. However, the value of  both quantities $d_B(M_R(s)\psi(s))$ and $d_R(M_B(s)\psi(s)$ defining the coupling {\it depend on what particular quantum state $\psi(s)$ we choose}. Thus, if we are interested in defining the continuous analog of $S_P(s)$ given in (\ref{Pageentropy}) we could start defining
\begin{equation}\label{algebraic1}
S^{alg}_P(s,\psi(s))=\min\{ d_B(M_R(s)\psi(s)),  d_R(M_B(s)\psi(s)\}
\end{equation}
where instead of using the finite dimensional $\log dH_{R,B}(s)$ we use the continuous dimensions $d_{B,R}(M_R(s)\psi(s))$. By contrast to the finite dimensional case where $dH_R(s)\cdot dH_B(s)=N$ for some finite and fixed $N$ defining the finite dimension of the Hilbert space now the only constraint comes from the independence on the state $\psi(s)$ of the ratio defining $d_B^{MvN}(s)$.

We can write (\ref{algebraic1}) in a more transparent way as
\begin{equation}\label{algebraic}
S^{alg}_P(s)=\min\{ \tau_{B(s)}(P_{R,s,\psi(s)}),  \tau_{R(s)}(P_{B,s,\psi(s)})\}
\end{equation}
with $\tau_B(s)$ and $\tau_R(s)$  the traces of the type $II_1$ factors $M_B(s)$ and $M_R(s)$ ( which are defined by the continuous dimensions $d_B$ and $d_R$ ) and with $P_{R,s,\psi(s)},P_{B,s,\psi(s)}$ the projections on the spaces $M_R(s)\psi(s)$ and $M_B(s)\psi(s)$. 

If we are interested in defining the continuous algebraic analog of Page's entropy $S_P(s)$ we need to fix a particular random state $\psi(s)$ and to define the corresponding average. For the enveloping Page curve $S_P(s)$ we choose, in the finite dimensional case, the state that maximizes  the entanglement entropy for the corresponding bipartite factorization. This corresponds to the {\it maximally entangled state}. Thus, what is the {\it continuous analog of such state} ?

\vspace{0.3 cm}

{\bf Step 4} Let us now address the question on how to define the continuous analog of the maximally entangled state of the bipartite case. In order to answer this question we need to recall some properties of the Murray von Neumann coupling. In particular:

\vspace{0.3 cm}

{\it If $d_B^{MvN}(s)$ is $\geq 1$ we can find a state $\psi(s)$ that is separating with respect to $M_B(s)$ while if $d_B^{MvN}(s)$ is $\leq 1$ we can find a state $\psi(s)$ that is cyclic with respect to $M_B(s)$.} 

\vspace{0.3 cm}

Thus only when $d_B^{MvN}(s)=1$ ( and consequently also $d_R^{MvN}(s)=1$ ) we can find a state $\psi(s)$ that is both {\it cyclic and separating}.

We can use this result to identify the particular state $\psi(s)$ playing the role of the {\it continuous} version of the {\it maximally entangled state} used in the definition of the enveloping Page curve.

In order to do it we will proceed as follows. Since the coupling is independent on the state $\psi(s)$  we can use the coupling to define different {\it phases} during the evaporation process. In particular we will define:

\vspace{0.3 cm}

{\it The $B$ phase if $d_B^{MvN}(s) \leq 1$ and the $R$ phase if $d_B^{MvN}(s) \geq 1$}

\vspace{0.3 cm}

Note that since $d_B^{MvN}(s)$ is the continuous version of the bipartite finite value of $\frac{d H_R(s)}{dH_B(s)}$ the regime $d_B^{MvN}(s) \leq 1$ corresponds to the early times of evaporation where we expect the dimension of the radiation Hilbert space to be smaller than the dimension of the black hole Hilbert space. Equivalently the $R$ phase is the continuous version of the regime of evaporation where, in the finite version, the radiation Hilbert space becomes larger than the black hole Hilbert space.

Now we can define the analog of the maximally entangled state used in Page's definition as follows:

\vspace{0.3 cm}

\begin{itemize}
\item  In the $B$ phase choose $\tilde \psi(s)$ cyclic w.r.t $M_B(s)$ i.e. separating w.r.t $M_R(s)$.
\item  In the $R$ phase choose $\tilde \psi(s)$ separating w.r.t $M_B(s)$ i.e. cyclic w.r.t $M_R(s)$.
\end{itemize}

\vspace{0.3 cm}

Note that the physics meaning of the state $\tilde \psi(s)$ is to be:

\vspace{0.3 cm}

{\it The state that is separating with respect to the algebra associated to what in the bipartite version would represents the smaller Hilbert space at that evaporation time.}

\vspace{0.3 cm}

In particular in the $B$ phase, where few quanta are radiated, $\tilde \psi(s)$ is separating with respect to the radiation algebra $M_R(s)$ while in the $R$ phase i.e. after the Page time, the state $\tilde \psi(s)$ is separating with respect to the black hole algebra $M_B(s)$. We will refer to this condition as {\it the separating condition}.

Once we have identified $\tilde \psi(s)$ as the state playing, along the evaporation process, the role of the maximally entangled state, we can define the algebraic version of Page's curve as
\begin{equation}\label{maximal}
S^{alg}_P(s)=\min\{ d_B(M_R(s)\tilde \psi(s)), d_R(M_B(s)\tilde \psi(s)\}
\end{equation}

\vspace{0.3 cm}

{\bf Remark} Note that the state $\tilde \psi(s)$ being separating with respect to $M_R(s)$ in the $B$ phase and separating with respect to $M_B(s)$ in the $R$ phase is not necessarily unique. Thus we could define the algebraic version of Page curve using some average on random states $\tilde \psi(s)$ or choosing the state $\tilde \psi(s)$ that maximizes (\ref{maximal}). 

\vspace{0.3 cm}

{\bf Step 5} Let us now focus our attention on how the information paradox could be solved in this algebraic setup. Recall that in the finite bipartite approach the information paradox was solved assuming that {\it $d_P(s)$, defined in (\ref{algebraic}), grows monotonically from $\frac{1}{N}$ to $N$ for $N$ the finite dimension of the full Hilbert space}. Page's time being identified by $d_P(s_P)=1$. The algebraic continuous version of this assumption is that:

\vspace{0.3 cm}

{\it The coupling $d_B^{MvN}(s)$ grows monotonically from a minimal value $\delta$ ( with $\delta<<1$ ) at the moment of black hole formation to some value $\frac{1}{\delta}$ at the end of evaporation.}

\vspace{0.3 cm}

Note that from this assumption it follows that the algebraic curve defined by (\ref{maximal}), for {\it any} state $\tilde \psi(s)$ satisfying the {\it separating condition}, follows a Page like curve. Note also that at this point we are not fixing the value of $\delta$.

Before justifying this assumption let us see what it is telling us about the nature of {\it the phase transition at Page's time}.

A monotonic growth of $d_B^{MvN}(s)$ implies that if we start in the $B$ phase with $d_B^{MvN}(s)<1$ for $s<s_P$ we move into the $R$ phase at $s_P$ defined by $d_B^{MvN}(s_P)=1$. Hence, {\it what happens at this transition} ?

From the former discussion on the properties of the coupling it follows that:

\vspace{0.3 cm}

{\bf Lemma:} {\it At Page's time the state $\tilde \psi(s)$ that is separating w.r.t $M_R(s)$ for $s<s_P$ becomes non separating, w.r.t $M_R(s)$ for $s>s_P$.}

\vspace{0.3 cm}

This lemma follows from the assumption on the monotonic growth of the coupling $d_B^{MvN}(s)$. The transition at $s=s_P$ from {\it separating} into {\it non separating} ( w.r.t the finite factor associated to the radiation entanglement wedge algebra $M_R(s)$ ) is at the core of the algebraic solution of the information paradox. Note that this is the exact algebraic continuous translation of what is taking place in the finite bipartite model used by Page. Indeed in the bipartite case with $H=H_1\otimes H_2$ the maximally entangled state can be written, using the Schmidt decomposition as
\begin{equation}
\psi=\sum_{i:1,\min\{dH_1,dH_2\}}c_i\psi_i\psi^{'}_i
\end{equation}
Now for all the $c_i\neq0$ the state $\psi$ will be for $dH_1<dH_2$ separating with respect $B(H_1)$ but not with respect to $B(H_2)$ while for $dH_2<dH_1$ the state will be separating for $B(H_2)$ but not for $B(H_1)$.

\vspace{0.3 cm}

{\bf Step 6} Let us now address the problem of identifying the algebraic continuous version of the information functions $I^{B,R}(s)$ defined in (\ref{information}). Recall that for $s>s_P$
\begin{equation}\label{infopage}
I^R(s)=\log d_P(s)
\end{equation}
To identify the continuous version let us rewrite (\ref{infopage}) as
\begin{equation}\label{infopage2}
I^R(s)= \log(\frac{dH_R(s) dH}{dH_B(s) dH})
\end{equation}
with $dH$ the dimension of the full Hilbert space. The continuous version of (\ref{infopage2}) where we use continuous dimensions should be defined as
\begin{equation}
I_{alg}^R(s)=\log\frac{d_B(M_R(s)\tilde \psi(s)) d_R(H_s)}{d_R(M_B(s)\tilde \psi(s)) d_B(H_s)}
\end{equation}
where $H_s$ is the Hilbert space on which $M_B(s)$ and $M_R(s)$ are acting. This quantity is the Murray von Neumann parameter $\theta^{MvN}(s)$ defined in \cite{MvN2}. This leads to 
\begin{equation}
I^R_{alg}(s)=\log \theta^{MvN}(s)
\end{equation}
In agreement with the result in \cite{Gomez1} on the algebraic representation of the {\it information transfer} \footnote{For a type $III$ approach to the transfer of information along evaporation see \cite{EL1}.}.

\section{Duality and the unitarity of the evaporation}
In the previous section we have identified, in the finite dimensional bipartite approach of Page, the condition for solving the information paradox i.e. for a unitary description of the full process of evaporation. This condition reduces to impose that $d_P(s)$, as defined in (\ref{algebraic}), grows monotonically from an initial value of $O(\frac{1}{N_{BH}})$, at the moment the black hole is formed, into a final value $O(N_{BH})$ at the end of the evaporation process. Here by $N_{BH}$ we mean the finite dimension of the full Hilbert space and by the parameter $s$ the {\it evaporation time} that we set to zero at the initial moment of black hole formation.

The continuous algebraic version \cite{Gomez1} of this {\it unitarity condition}, described in the previous section, corresponds to replace the finite dimensional parameter $d_P(s)$ by the Murray von Neumann coupling $d_B^{MvN}(s)$ where $d_B^{MvN}(s)$ is the coupling for the type $II_1$ factor $M_B(s)$ describing the black hole entanglement wedge at time $s$. Once we use the coupling instead of $d_P(s)$ the unitarity condition needed to avoid the information paradox becomes the monotonic growth of $d_B^{MvN}(s)$ from some initial value $\delta=d_B^{MvN}(s=0)$ with $\delta << 1$ into a final value $O(\frac{1}{\delta})$ at the end of evaporation. Here the value of $\delta$ defines {\it the algebraic initial conditions} associated to the moment of the black hole formation. That $\delta<<1$ reflects that at this time the black hole has not emitted any substantial radiation. In the continuous case and if we work with finite type $II_1$ factors we need to be careful to identify $\delta$ since in this case we don't have minimal projections. In the algebraic version the Page's time $s_P$ is determined by the condition $d_B^{MvN}(s_P)=1$ that is the continuous analog of the finite dimensional version $d_P(s_P)=1$.

In this general picture we are missing two basic ingredients, namely how to define the couple of type $II_1$ factors $(M_B(s),M_R(s))$ describing the evaporation process and what underlines the $s$ dependence of these factors in a way consistent with the {\it unitarity condition} i.e. with the monotonous growth of the coupling $d_B^{MvN}(s)$. In this section and following \cite{Gomez1} we will address both problems in a way as rigorous as possible.

\vspace{0.3 cm}

{\bf Step 1: The EPR model.} Let us start with the Penrose diagram of the eternal black hole in Figure 3. Using {\it duality} in the way described in chapter 2 we define the centralizer $(A_R\otimes F_{\infty})_{\omega}$. Recall that the type I factor $F_{\infty}$ emerges naturaly, thanks to Takesaki duality, once we implement covariantly the modular automorphisms of $A_R$ as well as its dual. Moreover the dominant weight is uniquely determined once we identify the action of modular automorphisms on $A_R\otimes F_{\infty}$. Thus the only ambiguity defining the centralizer $(A_R\otimes F_{\infty})_{\omega}$ is the weight $\phi$ on $A_R$ used to define the corresponding GNS representation of $A_R$, that we assume, following Araki's basic result \cite{Araki} is a type $III_1$ factor. The algebra $(A_R\otimes F_{\infty})_{\omega}$ is a type $II_{\infty}$ factor. As the first step we will define an EPR model of this factor. 

If $(A_R\otimes F_{\infty})_{\omega}$ is hyperfinite, as we will assume, this factor is isomorphic to the {\it matrix amplification} ${\cal{M}}(R)$ for $R$ the hyperfinite type $II_1$ factor defined by von Neumann using infinite pairs of EPR pairs. In Appendix A we will briefly review von Neumann's construction of $R$. The definition of the {\it matrix amplification} ${\cal{M}}(R)$ goes as follows. 

For any finite $n$ define the type $II_1$ factor $M_n(R)$ of $(n\times n)$ matrices valued in $R$ i.e. matrices $(a_{i,j})$ with $i,j=1...n$ and with $a_{i,j}$ elements in $R$. Denoting the trace on $R$ as $d_R$ the trace on $M_n(R)$ is naturally defined by $(Tr_n\otimes d_R)$ i.e.
\begin{equation}\label{trace}
Tr_n(a_{i,j})= \sum_{i=1..n} d_R(a_{i,i})
\end{equation}
The algebra ${\cal{M}}(R)$ is defined as
\begin{equation}
{\cal{M}}(R)=\cup_{n\leq 1} M_n(R)
\end{equation}
with the semifinite trace $Tr$ on ${\cal{M}}(R)$ naturally defined by $(Tr\otimes d_R)$. Once we model $(A_R\otimes F_{\infty})_{\omega}$ as ${\cal{M}}(R)$ we can identify the trace defined as $\tau_{\omega}$ with the trace on ${\cal{M}}(R)$ defined by $Tr\otimes d_R)$.

Let us now define $M_B(0)$ at the initial time as a type $II_1$ factor in ${\cal{M}}(R)$ that we will denote $R_{0}$.

\vspace{0.3 cm}

{\bf Step 2: Initial conditions.} Our next task will be to characterize $R_0$ i.e. the initial conditions of the evaporation process. The conditions defining $R_0$ are the following:

\vspace{0.3 cm}
 
i) There exist a natural number $n$ such that $R_0\in M_n(R)$

\vspace{0.3 cm}

ii) There exist a finite projection $P_0$ in the space of projections of $M_n(R)$ such that
\begin{equation}
R_0=P_0M_n(R)P_0
\end{equation}
and such that 
\begin{equation}
Tr(P_0)= \Delta_0
\end{equation}
with $\Delta_0$ some real number, at this point undetermined, defining the initial conditions.

\vspace{0.3 cm}

iii) The Murray von Neumann coupling $d_{R_0}^{MvN}=\delta$ with $\delta<<1$. Recall that this $\delta$ defines the initial conditions of black hole formation. However at this point is, as well as $\Delta_0$,  just {\it phenomenological data} defining the initial conditions.

\vspace{0.3 cm}

{\bf Step 3: Duality.} On ${\cal{M}}(R)$ we can define the {\it dual automorphisms} $\hat \sigma_s$ introduced in chapter 2. What we know about these dual automorphisms is that they scale the trace. More precisely let us define $P_0^s=\hat \sigma_s(P_0)$ then
\begin{equation}
Tr(P_0^s)= e^{-s}Tr(P_0)=e^{-s}\Delta_0
\end{equation}
Since our final goal is to identify the dual parameter $s$ with a way to parametrize the evaporation process let us associate to the dual modular automorphism $\hat\sigma_s$ the transformation
\begin{equation}
R_0\rightarrow R_0^s = P_0^s {\cal{M}}(R) P_0^s
\end{equation} 
and let us think of $R_0^s$ as the type $II_1$ factor $M_B(s)$ i.e. the one describing the black hole entanglement wedge at "evaporation time" $s$.

At this point it is convenient to use the notation introduced by Murray and von Neumann in \cite{MvN1} ( see Appendix A for details). In this notation for a generic type $II_1$ factor $M$ and a projection $P$ in the space of projections of ${\cal{M}}(M)$, let us say ${\cal{P}}({\cal{M}}(M))$ with $Tr(P)=t$ we define as $M^t$ the algebraic class of the factor $P{\cal{M}}(M)P$. Using this notation for our case ${\cal{M}}(R)$ we find
\begin{equation}
R_0=R^{\Delta_0}
\end{equation}
and
\begin{equation}\label{constituency1}
R_0^s= R^{e^{-s}\Delta_0}
\end{equation}

\vspace{0.3 cm}

{\bf Step 4: The unitarity condition.} At this point we will solve the information paradox i.e. we will achieve the algebraic unitarity condition if: 

\vspace{0.3 cm}

i) We identify the evaporation process with $M_B(s)=R_0^s$ with $s$ parametrizing the dual modular automorphism and

\vspace{0.3 cm}

ii) If we can prove that $d_{R_0^s}^{MvN}$ grows monotonically with $s$

\vspace{0.3 cm}

In order to prove the statement ii) let us evaluate the coupling $d_{R_0^s}^{MvN}$. Recall that we have defined the initial conditions of the evaporation process by $d_{R_0}^{MvN}=\delta$. Let us now look for a projection $p_s$ in the space of projections ${\cal{P}}(R_0)$ such that
\begin{equation}\label{condition}
R_0^s=p_sR_0p_s
\end{equation}
The condition (\ref{condition}) implies that $p_s$ should satisfy
\begin{equation}\label{condition1}
Tr(p_s)= e^{-s}
\end{equation}
Recall that in the initial conditions we have fixed $R_0$ to be in $M_n(R)$ for some finite $n$ that we have not fixed. The trace $Tr$ in (\ref{condition1}) is defined as $(Tr_n\otimes d_R)$ which is the trace $Tr$ introduced in (\ref{trace}). Since $p_s$ should be a projection in the type $II_1$ factor $R_0$ we need to limit the values of $s$ to the interval $[0,\infty]$ i.e. to ${\mathbb{R}}^+$ \footnote{Indeed $Tr(p)\in [0,1]$ for $p$ any projection in $R_0$.}. Note that a priori the dual modular automorphisms $\hat \sigma_s$ are defined for $s$ any real number. Thus we limit the evaporation process to the positive real line i.e. to positive values of $s$. 

In these conditions we can easily evaluate the value of the coupling $d_{R_0^s}^{MvN}$ obtaining \footnote{Where we use that $d^{MvN}_{p_sR_0p_s}=d^{MvN}_{R_0} \frac{1}{Tr(p_s)}$.}
\begin{equation}\label{evaporatioindex}
d_{R_0^s}^{MvN}= \delta e^{s}
\end{equation}
This result solves the {\it algebraic unitarity condition} and consequently the information paradox during the evaporation process with $s\in[0,\infty]$.

\vspace{0.3 cm}

{\bf Step 5: Fixing the initial conditions.} It remains now to fix the values $\Delta_0$ and $\delta$ defining the algebraic initial conditions. From (\ref{evaporatioindex}) it follows that the Page time $s_P$ is given by
\begin{equation}
e^{s_P}=\frac{1}{\delta}
\end{equation}

However since at $s=s_P$ we have that $d_{R_0^s}^{MvN}=1$ at this time $R_0^s$ should be algebraically identical to $R$ i.e. to the EPR hyperfinite type $II_1$ factor. This leads to the algebraic isomorphism
\begin{equation}
R_0^{s_P} = R
\end{equation}
that using (\ref{constituency1}) implies
\begin{equation}
\Delta_0 = \frac{1}{\delta}
\end{equation}
This reduces the algebraic phenomenological initial data to just the value of $\delta$. If we assume that at the end of evaporation $d_{R_0^{s_{end}}}=\frac{1}{\delta}$ we will get $e^{s_{end}} \sim e^{2s_P}$.

\vspace{0.3 cm}

{\bf Step 6: Hilbert space description}
Until this point we have described the evaporation process at a purely algebraic level identifying $M_B(s)$ as $R_0^s$. The factor $R_0$ is acting on the Hilbert space $P_0\hat H$ with $\hat H= L^2({\mathbb{R}},H_{EPR})$ and with $H_{EPR}$ the GNS Hilbert space representation of the hyperfinite factor $R$ described in Appendix A. Moreover the factor $R_0^s$ is acting on the Hilbert space $p_sP_0\hat H$ that we will denote $\hat{H}(s)$. At each time of the evaporation process we can define the vector state $\tilde \psi(s)\in \hat{H}(s)$ such that is, for $s<s_P$, separating w.r.t $M_R(s)$ and {\it non separating} for $s>s_P$.

Thus the evaporation process is described by a function $\tilde \psi(s): s\rightarrow \hat H$. Since the type $II_1$ factors $R_0^s$ describing the evaporation process are related by the action of the dual automorphism $\hat \sigma_s$ we can identify the state $\tilde \psi(s)$ as an element in the Hilbert space $\hat {\cal{H}}$ representing the crossed product $M(R)\rtimes_{\hat \sigma}{\mathbb{R}}$ i.e
\begin{equation}
\hat{\cal{H}}= L^2({\mathbb{R}},\hat{H})
\end{equation}
In this space we can define a {\it unitary} representation $\hat{V}_s$ of the dual automorphisms, namely the one defining the crossed product $M(R)\rtimes_{\hat \sigma}{\mathbb{R}}$, with
\begin{equation}
\hat{V}_{s^{'}} \tilde \psi(s) = \tilde \psi (s+s^{'})
\end{equation}
This makes $\hat V_s$ the {\it unitary implementation of the evaporation process} on $\hat{\cal{H}}$. Using now Takesaki duality and the isomorphism between $M(R)\rtimes_{\hat \sigma}{\mathbb{R}}$ and $A_R\otimes F_{\infty}$ with $A_R$ representing the type $III_1$ factor, associated with the black hole entanglement wedge at initial time, we can map $\hat{V}(s)$ into the Weyl dual $V_s$ of $U_t$ defining the {\it emergent observer} algebra $F_{\infty}$.

In summary:

\vspace{0.3 cm}

{\it The Hilbert space representation of the evaporation process is defined by a path $\tilde \psi(s)$  in the Hilbert space representation of the dual crossed product $M(R)\rtimes_{\hat \sigma}{\mathbb{R}}$.}

\vspace{0.3 cm}

Nicely enough the Hilbert space representation of the evaporation process can be carried out into the Hilbert space representing the original type $III_1$ factor $\otimes$ the emergent observer factor $F_{\infty}$. 

We can summarize the full construction in the following Lemma.
 
 \vspace{0.3 cm}
 
{\bf Lemma:} {\it Given the initial data $\delta$ defining the Murray von Neumann coupling at initial time the  unitary evaporation process is defined by the family of type $II_1$ factors $R_0^s =R^{\frac{e^{-s}}{\delta}}$ ( and their conmutants) with $R$ the hyperfinite EPR factor.}

\vspace{0.3 cm}
 
Moreover the dual automorphism implements unitarily the evaporation process.
 
 \vspace{0.3 cm}
 
{\bf Step 7: Energetics and $\hbar$}. 
The fundamental lesson we learned in section 2.7 was to identify the Weyl algebra defined by $U_t$ and $V_s$ as the root of the quantum nature of dual automorphisms. Once the algebraic approach to the evaporation process has been defined, the constant $\hbar$ defining the Weyl Heisenber algebra associated to the covariant implementation of $G\times \hat G$ emerges as the characteristic quantum parameter of the evaporation process. If it is zero, the black hole will become eternal. In section 2.4 we discussed how to define asymptotic energies using $H_R$ or $H_L$ that are well defined operators in the corresponding crossed products ${\cal{A}}_{R,L}$. In order to convert these divergent energies into something finite we use finite projections in ${\cal{A}}_{R}$ for example $P_0$ defining the initial conditions. Formally this should allows us to define in terms of $P(s)H_RP(s)$ an energy dependent on the evaporation time $s$ scaling with $s$ as $e^{-\hbar s}$ where we have reestablished ( in the sense discussed in 2.7)  the dependence on the quantization constant associated with the Weyl-Heisenberg representation of the duality. In the simplest approximation this indicates that the decay of the ADM mass of the black hole is proportional to this $\hbar$ with this $\hbar=0$ representing an {\it eternal black hole}.

\subsection{A final brief remark on Cosmology}
 As a concluding remark we could ask ourselves whether the duality, just discussed, has any relevance to {\it Inflationary Cosmology}. This question admits a more concrete formulation, namely: 
 
 {\it Can we use duality, in the sense described in this note, when defining the inflationary decay of a primordial form of dark energy?}
 
  If this were so, we would find that the slow roll parameter $\epsilon$ would be, as has been discussed, in a very preliminary manner in \cite{Gomez4}, reflecting what we have called here the duality $\hbar$. As discussed this $\hbar$ would define the quantumness of the {\it external clock} in the inflationary case. Duality defines a {\it Weyl clock algebra} but not a clock algebra with a well-defined bounded time operator. This leads to the discussion, started in \cite{Gomez4}, on how duality could be related to a clock model. These discussions go beyond the objective of this note and they enter into speculative regions that we will not touch in this note.

\begin{appendices}

\section{The EPR type $II_1$ factor}
In his seminal paper from 1939 \cite{vN1}, von Neumann constructed an explicit model of finite type $II_1$ factor in terms of an infinite number of EPR pairs of q-bits. This construction is enormously instructive in order to understand the physics meaning of von Neumann algebras and its deep connection with the notion of quantum entanglement. Although well known we have considered that it will be worth to highlight some of the key technical ingredients of this construction. Recall that it was in this construction where  the notion of crossed product was first introduced.
\subsection{Infinite direct products of q-bits}
The first ingredient of the construction is the definition of the Hilbert space representing an infinite, but numerable, set of q-bits. For a finite number $N$ of q-bits the Hilbert space is simply given by $H(N)=\otimes_{\alpha=1,,,N}H_{\alpha}$ with $H_{\alpha}$ the q-bit Hilbert space of dimension two. The first question addressed by von Neumann in \cite{vN1} was: 

\vspace{0.3 cm}

{\it How to define the $N=\infty$ limit of $H(N)$?} \footnote{Recall that in the bulk of this paper we were using the existence of particular $N=\infty$ limits of $I_{N}$ different from $I_{\infty}$. In this appendix we describe the original von Neumann construction of such type $II_1$ limits.}

\vspace{0.3 cm}

In order to answer this question let us introduce a numerable set
$I$ of labels i.e $\alpha\in I$ with $\alpha=1,2,....$ and let $H_{\alpha}$ be a collection of associated q-bit Hilbert spaces. We are interested in defining 
the infinite direct product $\otimes_{\alpha} H_{\alpha}$. In order to simplify the notation we will denote this space ${\cal{H}}(I)$. To construct this space we will start
considering
arbitrary sequences $(f_{\alpha})$ with each element in the sequence $f_{\alpha}\in H_{\alpha}$ and we will identify those infinite products $\otimes_{\alpha}f_{\alpha}$ satisfying the condition
\begin{equation}\label{one}
(\otimes_{\alpha} f_{\alpha},\otimes_{\alpha} g_{\alpha})= \prod_{\alpha}(f_{\alpha},g_{\alpha})
\end{equation}
where $(f_{\alpha},g_{\alpha})$ is the scalar product in $H_{\alpha}$. This condition implies that $||\otimes_{\alpha} f_{\alpha}|| = \prod_{\alpha} ||f_{\alpha}||$. Therefore elements $\otimes_{\alpha}f_{\alpha}$ satisfying (\ref{one}) should be associated with sequences $(f_{\alpha})$ such that $\prod_{\alpha} ||f_{\alpha}||$ is {\it convergent}. We will call these sequences $C$-sequences. 

\subsubsection{Quasi-convergence and quantum phases}
If we want to use $C$-sequences to define the infinite tensor product we need to require that for two C-sequences $(f_{\alpha})$ and $(g_{\alpha})$ the  
$\prod_{\alpha}(f_{\alpha},g_{\alpha})$ should be also convergent. This is not automatically the case when we include {\it quantum phases}. Indeed, imagine a $C$-sequence $(f_{\alpha})$ and another $C$-sequence $(g_{\alpha})$ defined as $z_{\alpha}f_{\alpha}$ with $z_{\alpha}=|z_{\alpha}|e^{i\theta_{\alpha}}$ and $|z_{\alpha}|=1$. In this case we have that although $\prod ||f_{\alpha}|| = \prod ||g_{\alpha}||$ and both are  convergent, however $\prod_{\alpha}(f_{\alpha},g_{\alpha})=\prod z_{\alpha}\prod(f_{\alpha},f_{\alpha})$ does not need to be convergent if the sum of the phases $\sum_{\alpha} \theta_{\alpha}$ is divergent. 

This phenomena leads to von Neumann to introduce the notion of {\it quasi convergency}. Generically, for an arbitrary sequence of complex numbers $z_{\alpha}$, this sequence is said to be {\it quasi convergent} iff $\prod |z_{\alpha}|$ is convergent {\it but not} the $\sum\theta_{\alpha}$. von Neumann definition of quasi convergency assigns to quasi convergent sequences $z_{\alpha}$ the value $\prod z_{\alpha}=0$. Note that with this definition of quasi convergency the $C$-sequences $f_{\alpha}$ and $g_{\alpha}= z_{\alpha}f_{\alpha}$ for $z_{\alpha}$ {\it quasi convergent} are orthogonal i.e. $(\otimes_{\alpha} f_{\alpha},\otimes_{\alpha} g_{\alpha}) = 0$. 

\subsubsection{Equivalent $C$-sequences}
We will say that a $C$-sequence is a $C_0$-sequence if $\otimes_{\alpha}f_{\alpha} \neq 0$. Now we will say that two $C_0$-sequences $f_{\alpha}$ and $g_{\alpha}$ are {\it equivalent} if they only differ i.e. $f_{\alpha}\neq g_{\alpha}$ in a {\it finite number} of components. For a given $C_0$-sequence $f_{\alpha}$ we will denote the corresponding {\it equivalence class} as $[f_{\alpha}]$. Note that two $C_0$-sequences in different equivalence classes are {\it orthogonal}
\begin{equation}
(\otimes f_{\alpha},\otimes g_{\alpha})=0
\end{equation}
In particular the $C_0$-sequences $(f_{\alpha})$ and $(z_{\alpha}f_{\alpha})$ are in the same equivalence class iff $\prod z_{\alpha}$ is convergent. By the definition of quasi convergency we get that the two $C_0$-sequences $f_{\alpha}$ and $z_{\alpha}f_{\alpha}$ are in different equivalent classes if $z_{\alpha}$ is quasi convergent ( but not convergent), namely in this case the definition of quasi convergency implies that $(\otimes f_{\alpha},\otimes z_{\alpha} f_{\alpha})=0$.

Once we have defined $C_0$- sequences our second step will consist in defining {\it finite} linear combinations of the type $\Psi=\sum_{\nu=1..p} (\otimes_{\alpha}f_{\alpha}^{\nu})$ and to  define the scalar product
\begin{equation}\label{scalar}
\langle \Psi,\Phi\rangle=\sum_{\nu,\mu} \langle \otimes_{\alpha}f_{\alpha}^{\nu},\otimes_{\alpha}g_{\alpha}^{\mu}\rangle
\end{equation}
for $\Phi=\sum_{\mu}\otimes_{\alpha} g_{\alpha}^{\mu}$ and with $ \langle \otimes_{\alpha}f_{\alpha}^{\nu},\otimes_{\alpha}g_{\alpha}^{\mu}\rangle=\prod_{\alpha}(f_{\alpha}^{\nu},g_{\alpha}^{\mu})$. The scalar product defined in (\ref{scalar}) allows us to define a {\it norm} as $||\Psi||^2=\langle\Psi,\Psi\rangle$ on the linear set of {\it finite linear combinations} of $C_0$ sequances. 

The final step will consist in defining the $N={\infty}$ limit of $H(N)$ that we will generically denote ${\cal{H}}(I)$, for $I$ the infinite set of q-bits, as:

\vspace{0.3 cm}

{\it The completion ,relative to the norm $||\Psi||^2=\langle\Psi,\Psi\rangle$, of the linear space of finite linear combinations of $C_0$ sequences.}

\vspace{0.3 cm}

\subsubsection{Decomposition of ${\cal{H}}(I)$ into subspaces}

For each equivalence class $[f_{\alpha}]$ we can define the associated {\it subspace} ${\cal{H}}([f_{\alpha}])$ of ${\cal{H}}(I)$ as the one generated by all finite linear combinations of $C_0$-sequences equivalent to $f_{\alpha}$. This defines a decomposition of 
${\cal{H}}(I)$ into {\it mutually orthogonal subspaces} each one associated with a different equivalence class. 

Let us denote $\Theta$ the abstract set representing the full set of equivalence classes and let us denote $[{\cal{A}}]$ each equivalence class in $\Theta$. Then the full set of equivalence classes $\Theta$ defines a complete decomposition of the von Neumann infinite product Hilbert space ${\cal{H}}(I)$ into mutually orthogonal subspaces. Contrary to ${\cal{H}}(I)$ that is {\it complete}, by construction, the subspaces ${\cal{H}}([\cal{A}])$ associated to each equivalence class are {\it not complete}.

\subsubsection{Weak equivalence and unitary evolution}
We can formally define associated with a sequence $z_{\alpha}$ of complex numbers with $|z_{\alpha}|=1$ a {\it unitary} operator \footnote{See Lemma 6.2.1 in \cite{vN1} for the proof of unitarity.} $U(z_{\alpha})$ transforming the $C$-sequences as 
\begin{equation}\label{unitarity}
U(z_{\alpha}) : f_{\alpha} \rightarrow z_{\alpha}f_{\alpha}
\end{equation}
Now we will define the {\it weak equivalence relation} as follows: two $C_0$-sequences $f_{\alpha}$ and $g_{\alpha}$ are weakly equivalent if there exist a sequence of complex numbers $z_{\alpha}$ such that $f_{\alpha}$ is equivalent to $z_{\alpha} g_{\alpha}$. 

As before we can use weak equivalence to define for any $C_0$-sequence $f_{\alpha}$ the weak equivalence class $[f_{\alpha}]_w$ \footnote{Note that $[f_{\alpha}]_w\supset [f_{\alpha}]$.}. The physics motivation for the definition of these weak equivalence classes is that they are preserved by the action of the unitary transformation $U(z_{\alpha})$ defined above. In other words {\it weak equivalence classes} define subspaces closed under the action of the unitary transformation defined by (\ref{unitarity}).

Note that for $z_{\alpha}$ quasi convergent the "evolution" operator $U(z_{\alpha})$ is defined on each {\it weak equivalence class} although it can transform one equivalence class into a different one.

\subsection{Associativity}
A key aspect of von Neumann  definition of infinite tensor products, crucial for the physics interpretation, is that they are not isomorphic under {\it associativity}. In particular we can decompose the infinite set of indices $I$ as $\cup_{\gamma}I_{\gamma}$ and to define, associated to this decomposition of $I$, the Hilbert space 
\begin{equation}
\otimes_{\gamma}{\cal{H}}(I_{\gamma})
\end{equation}
with ${\cal{H}}_{\gamma}=\otimes_{\alpha\in I_{\gamma}}H_{\alpha}$. Obviously for $I$ finite all the so defined spaces for different decompositions of $I$ are isomorphic and related by associativity. However, {\it this is not the case if the number of subsets $I_{\gamma}$ is infinite}. 

As an example imagine
you decompose the infinite set $I$ of q-bits into two equally infinite set of q-bits. Thus each q-bit can be labelled as $(n,\tau)$ with $n=1,2... $ and $\tau=0,1$. We can think the label $\tau=0,1$ as representing $L$ and $R$ q-bits. In other words we have the infinite set of $L$ q-bits $(n,L)$ and the equally infinite set of $R$ q-bits $(n,R)$. We will denote $H_{n,\tau}$ the two dimensional Hilbert space of each q-bit.

Now we can use naiv {\it associativity} to define different infinite tensor products. For instance ( see Figure 7)
( See Figure 7).
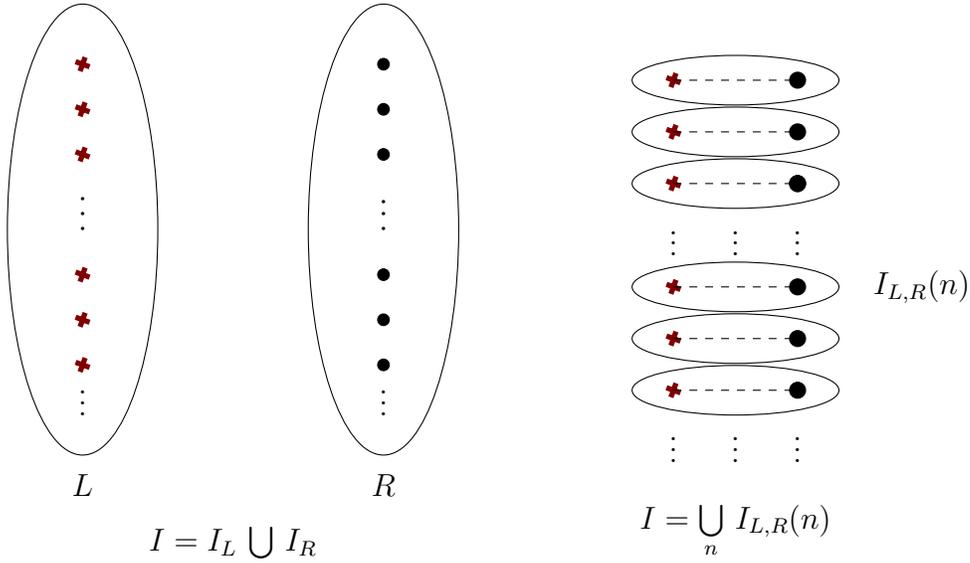
\begin{figure}
    \centering
    \begin{subfigure}[b]{0.45\textwidth}
        \centering
    \begin{tikzpicture}[scale=2]
        \draw (-1,0.1) circle(0pt) node[] {$\Huge\cdot$};
        \draw (-1,0) circle(0pt) node[] {$\Huge\cdot$};
        \draw (-1,-0.1) circle(0pt) node[] {$\Huge\cdot$};
        \draw (-1,1) node[cross=4pt] {};
        \draw (-1,0.7) node[cross=4pt] {};
        \draw (-1,-0.7) node[cross=4pt] {};
        \draw (-1,0.4) node[cross=4pt] {};;
        \draw (-1,-1) node[cross=4pt] {};
        \draw (-1,-0.4) node[cross=4pt] {};
        \draw (1,0.08) circle(0pt) node[] {$\Huge\cdot$};
        \draw (1,0) circle(0pt) node[] {$\Huge\cdot$};
        \draw (1,-0.1) circle(0pt) node[] {$\Huge\cdot$};
        \filldraw  (1,1) circle(1.1pt) {};
        \filldraw  (1,0.7) circle(1.1pt) {};
        \filldraw  (1,0.4) circle(1.1pt) {};
        \filldraw  (1,-1) circle(1.1pt) {};
        \filldraw  (1,-0.7) circle(1.1pt) {};
        \filldraw  (1,-0.4) circle(1.1pt) {};
        \draw[thick]  (-1,-1.65)circle(0pt) node[anchor=north ] {$L$};
        \draw[thick]  (1,-1.65)circle(0pt) node[anchor=north ] {$R$};
        \draw (-1,-.1) ellipse (.5cm and 1.5cm);
        \draw (1,-.1) ellipse (.5cm and 1.5cm);
        \draw[thick]  (0,-2)circle(0pt) node[anchor=north ] {$I=I_L\,\bigcup\, I_R$};
        \draw (1,-1.2) circle(0pt)node[ ] {$\vdots$};
        \draw (-1,-1.2) circle(0pt)node[ ] {$\vdots$};
    \end{tikzpicture}
\end{subfigure}
\begin{subfigure}[b]{0.45\textwidth}
    \centering
    \begin{tikzpicture}[scale=2.75]
        \draw (-.3,0) circle(0pt)node[ ] {$\vdots$};
        \draw (-.3,0.75) node[cross=4pt] {};
        \draw (-.3,0.5) node[cross=4pt] {};
        \draw (-.3,0.25) node[cross=4pt] {};
        \draw (-.3,-0.75) node[cross=4pt] {};
        \draw (-.3,-0.5) node[cross=4pt] {};
        \draw (-.3,-0.25) node[cross=4pt] {};
        \draw (.3,0) circle(0pt)node[ ] {$\vdots$};
        \filldraw  (0.3,0.25) circle(1.1pt) {};
        \draw (.9,-0.25) circle(0pt)node[ ] {$I_{L,R}(n)$};
        \filldraw  (.3,0.75) circle(1.1pt) {};
        \filldraw  (.3,0.5) circle(1.1pt) {};
        \filldraw  (.3,0.25) circle(1.1pt) {};
         \filldraw  (.3,-0.75) circle(1.1pt) {};
        \filldraw  (.3,-0.5) circle(1.1pt) {};
        \filldraw  (.3,-0.25) circle(1.1pt) {};
   \draw (0,0) circle(0pt)node[ ] {$\vdots$};

   \draw[dashed] (-.3,0.25) --(.3,.25);
   \draw[dashed] (-.3,0.5) --(.3,0.5);
   \draw[dashed] (-.3,0.75) --(.3,.75);
   \draw[dashed] (-.3,-0.25) --(.3,-.25);
   \draw[dashed] (-.3,-0.5) --(.3,-0.5);
   \draw[dashed] (-.3,-0.75) --(.3,-.75);
   \draw (0,0.25) ellipse (.5cm and .12cm);
   \draw (0,0.75) ellipse (.5cm and .12cm);
   \draw (0,-0.25) ellipse (.5cm and .12cm);
   \draw (0,0.5) ellipse (.5cm and .12cm);
   \draw (0,-0.5) ellipse (.5cm and .12cm);
   \draw (0,-0.75) ellipse (.5cm and .12cm);
   \draw (0,-1.25) circle(0pt)node[anchor=north ] {$I=\bigcup\limits_n\, I_{L,R}(n)$};
   \draw (0,-1) circle(0pt)node[ ] {$\vdots$};
   \draw (.3,-1) circle(0pt)node[ ] {$\vdots$};
   \draw (-.3,-1) circle(0pt)node[ ] {$\vdots$};    
\end{tikzpicture}
\end{subfigure}
    \caption{Two different decompositions of $I$ into subsets}\label{fig:7} 
\end{figure}

\vspace{0.3 cm}

i) $(\otimes_{n} H_{n,L})\otimes (\otimes_{n} H_{n,R})$ corresponding to decompose $I=I_L\cup I_R$ i.e to the case where $\gamma$ takes only two values $L$ and $R$. 

\vspace{0.3 cm}

ii)$\otimes_{n}(H_{n,L}\otimes H_{n,R})$. This corresponds to decompose $I=\cup_{\gamma} I_{\gamma}$ with $\gamma=1,2..n..$ the infinite set of natural numbers and each $I_{\gamma}$ the finite set composed of two elements $(n,L)$ and $(n,R)$.

\vspace{0.3 cm}

The second tensor product represents an infinite set of pairs of q-bits.  In case we work with a {\it finite} number of q-bits i.e. $n=1,2....N$ for some finite $N$ both Hilbert spaces are obviously isomorphic. Thus, for finite number of q-bits, we get associativity. However:

\vspace{0.3 cm}

{\it Is this the case when we deal with an infinite number of q-bits ? }

\vspace{0.3 cm}

The answer is negative as it was stressed by von Neumann in \cite{vN1}. To see this phenomena consider first the following simple example. Take two $C_0$-sequences in $\otimes_{n,\tau} H_{n,\tau}$, namely $(f_{\alpha})$ and $(g_{\alpha})$ defined by $g_{\alpha}=- f_{\alpha}$. Clearly these two sequences are {\it not equivalent} in $\otimes_{n,\tau} H_{n,\tau}$\footnote{The two sequences obviously differ in an infinite number of components.}. Let us now consider instead the space $\otimes_{n}(H_{n,L}\otimes H_{n,R})$. In this case the corresponding sequences are $(f_{n,L}\otimes f_{n,R})$ and $(g_{n,L}\otimes g_{n,R})=(-f_{n,L}\otimes -f_{n,R})$ which are clearly the same sequence and consequently equivalent. In other words:

\vspace{0.3 cm}

 {\it The set of equivalence classes, for infinite direct products, are not invariant under associativity}.

\vspace{0.3 cm}

Thus the set of equivalence classes of $\otimes_{n,\tau} H_{n,\tau}$ and of  $\otimes_{n}(H_{n,L}\otimes H_{n,R})$ are different and consequently in this case associativity is not defining an isomorphism at the level of equivalence classes.

This is the first lesson we extract from von Neumann definition of {\it infinite direct products}, namely that {\it associativity is not always defining an isomorphism that preserves the equivalence classes}. This is the case for the two Hilbert spaces defined above. Indeed, while the {\it "unentangled"} version $(\otimes_{n} H_{n,L})\otimes (\otimes_{n} H_{n,R})$ satisfies associativity the {\it "entangled"} version $\otimes_{n}(H_{n,L}\otimes H_{n,R})$ does not.

\subsubsection{Rings of operators: von Neumann algebras}
Now we are interested in defining algebras of bounded operators on the so defined infinite tensor products. For simplicity consider the case where all $H_{\alpha}$ are the two dimensional Hilbert space defining a q-bit. Take one particular arbitrary q-bit labelled by $\alpha_0$ and consider the full algebra of bounded operators $a_{\alpha_0}$ acting on $H_{\alpha_0}$. Note that for $H_{\alpha_0}$ the two dimensional Hilbert space representing one q-bit, this algebra is generated by some representation of Pauli matrices. We will refer to this algebra as $I_2=: M_2(\mathbb{C})$ with $M_2(\mathbb{C})$ the algebra of two by two complex matrices. Next we extend algebraically the algebra $I_2$ to act on $\otimes_{\alpha}H_{\alpha}$ by
\begin{equation}\label{extension}
\hat a_{\alpha_0}( \otimes f_{\alpha}) = a_{\alpha_0}f_{\alpha_0} \otimes (\otimes_{\alpha\neq \alpha_0} f_{\alpha})
\end{equation}
We will denote the set of all $\hat a_{\alpha_0}$ as ${\cal{B}}_{\alpha_0}$ and the ring of operators $\hat {\cal{B}}$ as the ring generated by ${\cal{B}}_{\alpha_0}$ for all $\alpha_0$ \footnote{These ring of operators in the sense used by Murray and von Neumann are what we know as von Neumann algebras, thus we will refer to them, from now on, as von Neumann algebras. See 5.2 of \cite{vN1}.}. 

Note that if we are considering a finite number $N$ of q-bits the ring $\hat {\cal{B}}$ is isomorphic to the space of bounded operators on the finite product $\otimes H_{\alpha}$ for $\alpha=1,2...N$. That this is not the case when we work with {\it infinite} tensor products is at the very core of the definition and deep meaning of von Neumann algebras. 

In order to understand this phenomena the following basic result on $\hat{\cal{B}}$ will be important. 
 
For a given $C_0$-sequence $f_{\alpha}^0$ let us consider the corresponding equivalence class $[f_{\alpha}^0]$ and the corresponding subspace $H([f_{\alpha}^0])$ generated by all $C_0$-sequences equivalent to $f_{\alpha}^0$ \footnote{Recall that this subspace is not complete.}. Let us denote this equivalence class $[{\cal{A}}]$. The von Neumann algebra $\hat{\cal{B}}$ defined above is acting on each equivalence class. Thus the reduction of $\hat{\cal{B}}$ to $H([f_{\alpha}^0])$ defines a von Neumann algebra acting on $H([f_{\alpha}^0])$. The reason is that any $a\in \hat{\cal{B}}$ commutes with the projections $P_{H([f_{\alpha}^0])}$ for any equivalence class. Thus for each equivalence class $[{\cal{A}}]$ we can reduce the action of $\hat{\cal{B}}$ on the associated subspace $H([{\cal{A}}])$. This defines a von Neumann algebra $\hat{\cal{B}}([\cal{A}])$ acting on $H([{\cal{A}}])$\footnote{This follows from Theorem X of \cite{vN1}.}.

The reason that explains why von Neumann algebras {\it know} about {\it entanglement} is now very clear. Namely von Neumann algebras are defined by reducing the action of the algebraically extended bounded operators acting on one q-bit ( i.e. on one $H_{\alpha}$ ) to the equivalence classes defining subspaces of the von Neumann infinite tensor product. But, as stressed, neither these equivalence classes nor the corresponding algebras $\hat{\cal{B}}([\cal{A}])$ are invariant under associativity in case we work with infinite number of q-bits. Note that for a {\it finite} number $N$ of q-bits we only have one equivalence class and consequently the corresponding von Neumann algebra $\hat{\cal{B}}([\cal{A}])$  is a {\it finite} type $I_N$ factor.

\subsubsection{The effect of associativity on the definition of von Neumann algebras}
Let us now make the former general comments more concrete using some examples. Imagine that you decompose the infinite number of q-bits into two infinite groups to be denoted, as before, $L$ and $R$. This means that you have two dimensional Hilbert spaces $H_{n,\tau}$ with $n=1,2,...$ and $\tau= L,R$. Now define, as we did before, the following two infinite product Hilbert spaces, namely:
\begin{equation}\label{noentangled}
(\otimes_{n} H_{n,L})\otimes (\otimes_{n} H_{n,R})
\end{equation}
and
\begin{equation}\label{entangled}
\otimes_{n} (H_{n,L}\otimes H_{n,R})
\end{equation}

Let us now take the algebra ${\cal{B}}^{L}_{n_0}$ of bounded operators acting on one $L$ q-bit let us say $H_{n_0,L}$. Now you want to extend {\it algebraically} ${\cal{B}}^{L}_{n_0}$ to act on the Hilbert spaces (\ref{noentangled}) and (\ref{entangled}). As we will see this algebraic extension will produce {\it different von Neumann algebras on the corresponding different equivalence classes}.

In the case you consider the Hilbert space (\ref{noentangled}) you will define the algebraic extension as follows. Take an operator $a$ in ${\cal{B}}_{n_0}(L)$ and extended it first to $(\otimes_{n} H_{n,L})$ using (\ref{extension}). Let us denote the set of these operators $\hat {\cal{B}}^{L}_{n_0}$. Let us denote $\hat {\cal{B}}^{L}$ the ring generated by all the $\hat {\cal{B}}^{L}_{n_0}$ with $n_0=1,2,...$. Next you will extend $\hat {\cal{B}}^{L}$ to $(\otimes_{n} H_{n,L})\otimes (\otimes_{n} H_{n,R})$. Obviously this extension will produce a ring isomorphic to $\hat{\cal{B}}^{L}$ i.e. the ring of bounded operators acting on $\otimes_{n} H_{n,L}$. In this case we see that:

\vspace{0.3 cm}

$\hat{\cal{B}}^{L}$ is a type $I_{\infty}$ factor

\vspace{0.3 cm}

In other words for the Hilbert space (\ref{noentangled}) we get a type $I_{\infty}$ factor.

Let us now repeat the algebraic extension defining $\hat {\cal{B}}^{L}$ {\it but for the Hilbert space defined by (\ref{entangled})}. In other words we are trying to see {\it how associativity affects the definition of the algebra $\hat{\cal{B}}^{L}$}. For the Hilbert space (\ref{entangled}) the way to extend an operator $a$ acting on the Hilbert space $H_{n_0,L}$ of one q-bit will be:

\vspace{0.3 cm}

i) First extend it to $(H_{(n_0,L)}\otimes H_{(n_0,R)})$ and 

\vspace{0.3 cm}

ii) To extend it to the infinite product defined in (\ref{entangled}). 

\vspace{0.3 cm}

As before and using this algebraic extension defined by (\ref{entangled}) we can define the corresponding ring $\hat{\cal{B}}^{L}$. Let us now consider this algebra with a bit more detail.

\subsection{The hyperfinite EPR factor}
We will consider the two infinite families $L$ and $R$ of q-bits and the infinite product Hilbert space (\ref{entangled}). We will use the familiar physics notation with basis of each $H_{n,\tau}$ with $\tau=L,R$ given by $|+,L(R)\rangle_{n}$, $|-,L(R)\rangle_{n}$. We will define an equivalence class $[\alpha_n]$ as the set of $C_0$-sequences equivalent to the sequence defined by
\begin{equation}\label{reference}
g^0_n= \sqrt{\frac{1+\alpha_n}{2}}|+,L\rangle_n|+,R\rangle_n + \sqrt{\frac{1-\alpha_n}{2}}|-,L\rangle_n|-,R\rangle_n
\end{equation}
Let us consider the  simplest case, worked out by von Neumann \cite{vN1} and generalized by Powers \cite{Powers} and Araki and Woods \cite{AW}, with all $\alpha_n=\alpha$ equal and let us define
\begin{equation}
x= \frac{1-\alpha}{1+\alpha}
\end{equation}
It is easy to see that equivalence classes for different values of $x\in[0,1]$ are inequivalent\footnote{Note that two sequences $\alpha_n$ and $\alpha'_n$ are equivalent if $\sum_n(\sqrt{\frac{1-\alpha_n}{1+\alpha_n}}-\sqrt{\frac{1-\alpha'_n}{1+\alpha'_n}})^2$ is convergent. Thus two sequences with different values of $x$ are clearly inequivalent.}. Thus we will define the equivalence class as $[x]$. In Figure 8 we represent the reference state (\ref{reference}) defining the equivalence class $[x=1]$ and the equivalence class $[x=0]$. 

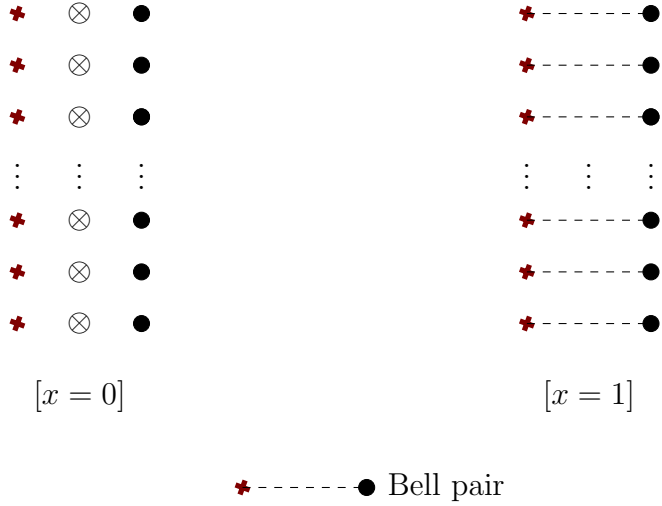
\begin{figure}
    \centering
    \begin{subfigure}[b]{0.4\textwidth}
        \centering
    \begin{tikzpicture}[scale=2.75]
        \draw (-.3,0) circle(0pt)node[ ] {$\vdots$};
        \draw (-.3,0.75) node[cross=4pt] {};
        \draw (-.3,0.5) node[cross=4pt] {};
        \draw (-.3,0.25) node[cross=4pt] {};
        \draw (-.3,-0.75) node[cross=4pt] {};
        \draw (-.3,-0.5) node[cross=4pt] {};
        \draw (-.3,-0.25) node[cross=4pt] {};
        \draw (.3,0) circle(0pt)node[ ] {$\vdots$};
        \filldraw  (0.3,0.25) circle(1.1pt) {};
        \filldraw  (.3,0.75) circle(1.1pt) {};
        \filldraw  (.3,0.5) circle(1.1pt) {};
        \filldraw  (.3,0.25) circle(1.1pt) {};
         \filldraw  (.3,-0.75) circle(1.1pt) {};
        \filldraw  (.3,-0.5) circle(1.1pt) {};
        \filldraw  (.3,-0.25) circle(1.1pt) {};
   \draw (0,0) circle(0pt)node[ ] {$\vdots$};
        \draw (0,0.75) circle(0pt)node[ ] {$\otimes$};
        \draw (0,.25) circle(0pt)node[ ] {$\otimes$};
        \draw (0,0.5) circle(0pt)node[ ] {$\otimes$};
        \draw (0,-0.5) circle(0pt)node[ ] {$\otimes$};
        \draw (0,-0.75) circle(0pt)node[ ] {$\otimes$};
        \draw (0,-0.25) circle(0pt)node[ ] {$\otimes$};
        \draw (0,-1.1) circle(0pt)node[ ] {$[x=0]$};
    \end{tikzpicture}
\end{subfigure}
    \begin{subfigure}[b]{0.4\textwidth}
        \centering
    \begin{tikzpicture}[scale=2.75]
        \draw (-.3,0) circle(0pt)node[ ] {$\vdots$};
        \draw (-.3,0.75) node[cross=4pt] {};
        \draw (-.3,0.5) node[cross=4pt] {};
        \draw (-.3,0.25) node[cross=4pt] {};
        \draw (-.3,-0.75) node[cross=4pt] {};
        \draw (-.3,-0.5) node[cross=4pt] {};
        \draw (-.3,-0.25) node[cross=4pt] {};
        \draw (.3,0) circle(0pt)node[ ] {$\vdots$};
        \filldraw  (0.3,0.25) circle(1.1pt) {};
        \filldraw  (.3,0.75) circle(1.1pt) {};
        \filldraw  (.3,0.5) circle(1.1pt) {};
        \filldraw  (.3,0.25) circle(1.1pt) {};
         \filldraw  (.3,-0.75) circle(1.1pt) {};
        \filldraw  (.3,-0.5) circle(1.1pt) {};
        \filldraw  (.3,-0.25) circle(1.1pt) {};
   \draw (0,0) circle(0pt)node[ ] {$\vdots$};

   \draw[dashed] (-.3,0.25) --(.3,.25);
   \draw[dashed] (-.3,0.5) --(.3,0.5);
   \draw[dashed] (-.3,0.75) --(.3,.75);
   \draw[dashed] (-.3,-0.25) --(.3,-.25);
   \draw[dashed] (-.3,-0.5) --(.3,-0.5);
   \draw[dashed] (-.3,-0.75) --(.3,-.75);
  
        \draw (0,-1.1) circle(0pt)node[ ] {$[x=1]$};
    \end{tikzpicture}
\end{subfigure}
\newline
\newline
\begin{subfigure}[b]{0.9\textwidth}
    \centering
\begin{tikzpicture}[scale=2.75]
    \draw (-.5,-0.25) node[cross=4pt] {};
   
    \filldraw  (.1,-0.25) circle(1.1pt) {};

\draw[dashed] (-.5,-0.25) --(.1,-.25);
    \draw (0.15,-0.25) circle(0pt)node[ anchor=west] {$\mbox{Bell pair}$};
\end{tikzpicture}
\end{subfigure}
\caption{Representation of the reference state ladders defining the equivalence classes $[x=0]$ and $[x=1]$. The associated von Neumann algebras are type $I_{\infty}$ for $[x=0]$ and type $II_1$ for $[x=1]$. The Hilbert spaces $H([x=0])$ and $H([x=1])$ are linearly generated by ladders with a final number of steps differing from the steps defining the reference state. Note that states in $H([x=0])$ represent states with finite {\it quantum entanglement} while states in $H([x=1])$ represent finite quantum fluctuations with respect to an {\it infinitely entangled reference state}.  }\label{fig:8} 

\end{figure}

Using the construction in the previous section we can define the algebras of operators $\hat{\cal{B}}^{L}([x])$ defined on the equivalence class $[x]$. These are the von Neumann-Powers factors $R_{[x]}$. Since we are defining these factors extending algebraically operators acting on one $L$ q-bit we can denote this factor $R^L_{[x]}$. Similarly we can define $R^R_{[x]}$. For this particular example the two constructions are perfectly symmetric and we can ignore the labels $L$ and $R$. 

Let us first consider the case $[x=0]$ corresponding to $\alpha=1$. For this value the reference state is an {\it unentangled product state} and the factor $R_{[0]}$ is  acting on the Hilbert space $H([x=0])$ and is a factor of type $I_{\infty}$. 

Next let us consider the case $[x=1]$ in this case the state (\ref{reference}) defining the equivalence class is an infinite product of EPR Bell states. Thus we will refer to $R_[1]$ as the EPR factor. These factors are hyperfinite and of type $II_1$. The Powers factors $R_[x]$ with $x\in]0,1[$ are inequivalent type $III$ factors.

\subsubsection{Bell states}
Following von Neumann in \cite{vN1} we will use the four Bell states to define a basis for the equivalence class $[x=1]$. For each couple of q-bits we can define the basis of $H_{n,L}\otimes H_{n,R}$ using the four Bell states that we will denote $f_{n,(\alpha,\beta)}$ with $\alpha,\beta = 0,1$. We will identify the Bell state $(0,0)$ with the state defined in (\ref{reference}) for $\alpha_n=0$ i.e. with the state used to define the equivalence class $[x=1]$. Use now the Pauli matrices $\sigma_{z,n}^L$ and $\sigma_{x,n}^L$ acting on the one q-bit basis $|\pm,L\rangle_n$ in the standard way, namely:
\begin{equation}
\sigma_{z,n}^L|\pm,L\rangle_n=\pm|\pm,L\rangle_n
\end{equation}
and
\begin{equation}
\sigma_{x,n}^L|\pm,L\rangle_n= |\mp,L\rangle_n
\end{equation}
The action on Bell states defined by algebraic extension is given by
\begin{equation}\label{bell1}
\sigma_{z,n}^L f_{n,(\alpha,\beta)} = (-1)^{\alpha}f_{n,(\alpha,1-\beta)}
\end{equation}
and
\begin{equation}\label{bell2}
\sigma_{x,n}^L f_{n,(\alpha,\beta)} = f_{n,(1-\alpha,\beta)}
\end{equation}
Now consider a basis of the EPR equivalence class $[x=1]$ defined by $\Phi_{\alpha_1,\beta_1,\alpha_2,\beta_2, ....}$ with $(\alpha_i,\beta_i) \neq (0,0)$ for only a {\it finite number of components}. Now we extend the action of $\sigma_{z,n}^L$ and $\sigma_{x,n}^L$ to the whole equivalence class as
\begin{equation}\label{generator}
\hat \sigma_{z,n}^L \Phi_{\alpha_1,\beta_1,...}=(-1)^{\alpha_n}\Phi_{\alpha_1,\beta_1,..\alpha_n,(1-\beta_n),...}
\end{equation}
and
\begin{equation}\label{generatorr}
\hat \sigma_{x,n}^L \Phi_{\alpha_1,\beta_1,...}= \Phi_{\alpha_1,\beta_1,..,(1-\alpha_n),\beta_n,...}
\end{equation}

The algebra generated by these operators is the EPR type $II_1$ factor defined by von Neumann in \cite{vN1} \cite{MvN2}. We will denote this algebra $\hat{\cal{B}}^L_{EPR}$ or more simply ( as we did in section 4)  as $R$. \footnote{Notice that the same construction using the $R$ q-bits leads to the definition of an isomorphic factor.}.

This factor is a typical example of the so called ITPFI ( infinite tensor products of type $I$ factors). In the case of $R$ we are using the $I_2$ algebra of bounded operators acting on the q-bit two dimensional Hilbert space i.e. the Pauli matrices. Now we are defining the infinite tensor product as $\otimes_n(H_{n,L}\otimes H_{n,R})$ with $I_2$ acting on the four dimensional Hilbert space $(H_{n,L}\otimes H_{n,R})$ in the form defined by (\ref{bell1}) and (\ref{bell2}). The action of $R$ on the Hilbert space $H([x=1])$ just defined is an example of the GNS construction.

\subsubsection{Araki-Woods factor}

A particularly interesting factor can be defined using and infinite family of $L$ and $R$ {\it qtrits}. In this case $H_n=H_{n,L}\otimes H_{n,R}$ will be six dimensional and $H_{n,L}$ and $H_{n,R}$ will represent the three dimensional Hilbert spaces of qtrits. Now we can extend algebraically the action of $I_3$ (i.e. of $M_3(\mathbb{C})$) to $H_{n,L}\otimes H_{n,R}$ as $I_3\otimes 1$. We can choose the state $\Omega_n$ in $H_n$ as the following analog of a Bell state for qtrits.
Taking as the qtrit basis $|0\rangle,|1\rangle,|2\rangle$ we define
\begin{equation}
\Omega_n=\frac{1}{\sqrt{1+\lambda+\mu)}}(|0,L\rangle \otimes |0,R\rangle+\sqrt{\lambda}|1,L\rangle \otimes |1,R\rangle +\sqrt{\mu}|2,L\rangle \otimes |2,R\rangle)
\end{equation}
with $\lambda$ and $\mu$ independent of $n$. The ITPFI factor defined as $\otimes (H_n,\Omega_n)$ is acting on the corresponding equivalence class $[\Omega_n]$. In case 
\begin{equation}
\frac{\log \lambda}{\log \mu}
\end{equation}
is {\it irrational} the so defined IFPFI is the Araki Wodds $R_{\infty}$ factor that is a type $III_1$ factor. Hence we can define the representation
\begin{equation}
R_{\infty}=R_{\lambda}\otimes R_{\mu}
\end{equation}
for $\lambda$ and $\mu$ such that $\frac{\log \lambda}{\log \mu}\notin{\mathbb{Q}}$. Another useful representation of $R_{\infty}$ is $R_{\infty}=\otimes_{n=2}^{\infty} R_{\frac{1}{n}}$. 

\vspace{0.3 cm}

In this appendix we will not discuss neither this type $III_1$ factor nor the important result of Connes and Haagerup \cite{haagerup} on the uniqueness and equivalence of  hyperfinite type $III_1$ factors to the Araki Woods $R_{\infty}$ factor.

\subsection{Matrix amplifications}
Given the EPR factor $R$ acting on $H_{EPR}$ ( defined as the Hilbert space associated with the equivalence class $[x=1]$ ) we can define for any integer $n$ an amplification $M_n(R_{EPR})$ as {\it the algebra of $n\times n$ matrices with entries in $R$}. Thus this algebra is isomorphic to $R\otimes M_n({\mathbb{C}})$ with $M_n({\mathbb{C}})$ the algebra of $n\times n$ complex matrices. The algebra $M_n(R)$ is acting on $H_{EPR}^{\oplus n} = {\mathbb{C}}^n\otimes H_{EPR}$. All these algebras $M_n(R)$ {\it are different type $II_1$ factors}. 

Now we can define a type $II_{\infty}$ EPR factor ${\cal{M}}(R)$ as
\begin{equation}
{\cal{M}}(R) =\bigcup_n M_n(R)
\end{equation}
for $n \geq 1$. This type $II_{\infty}$ factor is acting on the Hilbert space
\begin{equation}\label{Hilbert}
{\cal{H}}=H_{EPR}\otimes l^2(\mathbb{N})
\end{equation}
These are the factors used in section 4.

In general we can define a matrix amplification of any finite type $II_1$ factor $M$ using a set $I$ of {\it species} and defining the amplification using matrix $(m)_{i,j}$ valued in $M$. When the set $I$ of species is infinite and numerable we get a type $II_{\infty}$ factor.
 On the type $II_1$ factor $R$ we can define a trace $\tau_{EPR}$  using the {\it continuous dimension $d_R$}. This trace can be normalized and maps the set of equivalence classes of projections $p\in R$ into $[0,1]$. That for any $t\in[0,1]$ we can find a projection $p\in R$ such that $\tau_{EPR}(p)=t$ follows from the existence, for any finite natural number $n$, of projections $p_n\in R$ such that $\tau_{EPR}(p_n)=\frac{1}{2^n}$ and from the fact that we can find a set of projections $p_{n_{i}}$ mutually orthogonal and such that $\tau_{EPR} (\sum_i p_{n_i}) = \sum_i \frac{1}{2^{n_i}}$ defines the dyadic decomposition $t=\sum_i\frac{1}{2^{n_i}}$. In this case the projection $p$ with $\tau_{EPR}(p)=t$ is $p=\sum_{i} p_{n_i}$ \footnote{The former result is generic for any type $II_1$ factor. Indeed since for a type $II_1$ factor $M$ does not exist a minimal projection we can find for any natural number $n$ a projection $p_n$ such that $\tau_M(p_n)=\frac{1}{2^n}$. Consequently for any $p$ with $\tau_M(p)=t$ we can find, using the dyadic expansion of $t$, the set of mutually orthogonal projections $p_{n_i}$ such that $p=\sum_i p_{n_i}$.}

In order to define a semi finite trace $Tr$ on the type $II_{\infty}$ matrix amplification ${\cal{M}}(R)$ we use the matrix representation of any element $x\in {\cal{M}}(R)$ as an infinite matrix with entries $(x)_{i,j}$ in $R$ and define $Tr(x)=\sum_{i}\tau_{M}(x_{i,i})$. This trace cannot be normalized and defines a weight on ${\cal{M}}^{+}(R)$.

Now consider a projection $P\in {\cal{M}}(R)$. We say that this projection is finite if $Tr(P)$ is finite. Note that if $Tr(P)$ is finite and equal to $s$ where $s$ now can be any finite positive real number in $[0,\infty]$ there exist a natural number $n$ such that $P\in M_n(R_{EPR})$. Moreover $P{\cal{M}}(R)P$ is isomorphic to the type $II_1$ factor $PM_n(R)P$. In other words for any finite projection $P_s$ with $Tr(P_s)=s$ we can define the type $II_1$ factor
\begin{equation}
M^s(R)= P_s{\cal{M}}(R)P_s
\end{equation}

The type $II_1$ factor $M^s(R)$ is acting on $P_s{\cal{H}}$ where ${\cal{H}}$ is defined in (\ref{Hilbert}). In the text we have denoted $M^s(R)$ simply as $R^s$.

\subsection{$q$-bit Holography: an algebraic approach to the EPR/ER relation}
 Let us now define, on the basis of von Neumann's definition of infinite tensor products, a $q$-bit holographic kit ( see Figure 9). Let us define as the analog of the {\it boundary CFT Hilbert space} the infinite tensor product of an infinite set $I$ of q-bits. For any partition $I=\cup_{\gamma}I_{\gamma}$ we define the corresponding Hilbert space $H(I)$. Recall that different decompositions related by associativity don't lead, in general, to isomorphic Hilbert spaces. Now we will define the correspondence:
 
 \vspace{0.3 cm}
 
 {\it To each equivalence class $[\alpha]$ in $H(I)$ we associate a {\it bulk geometry $g([\alpha])$} with $H([\alpha])$ representing the Hilbert space of quantum fluctuations around the geometry $g([\alpha])$.}
 
 \vspace{0.3 cm}

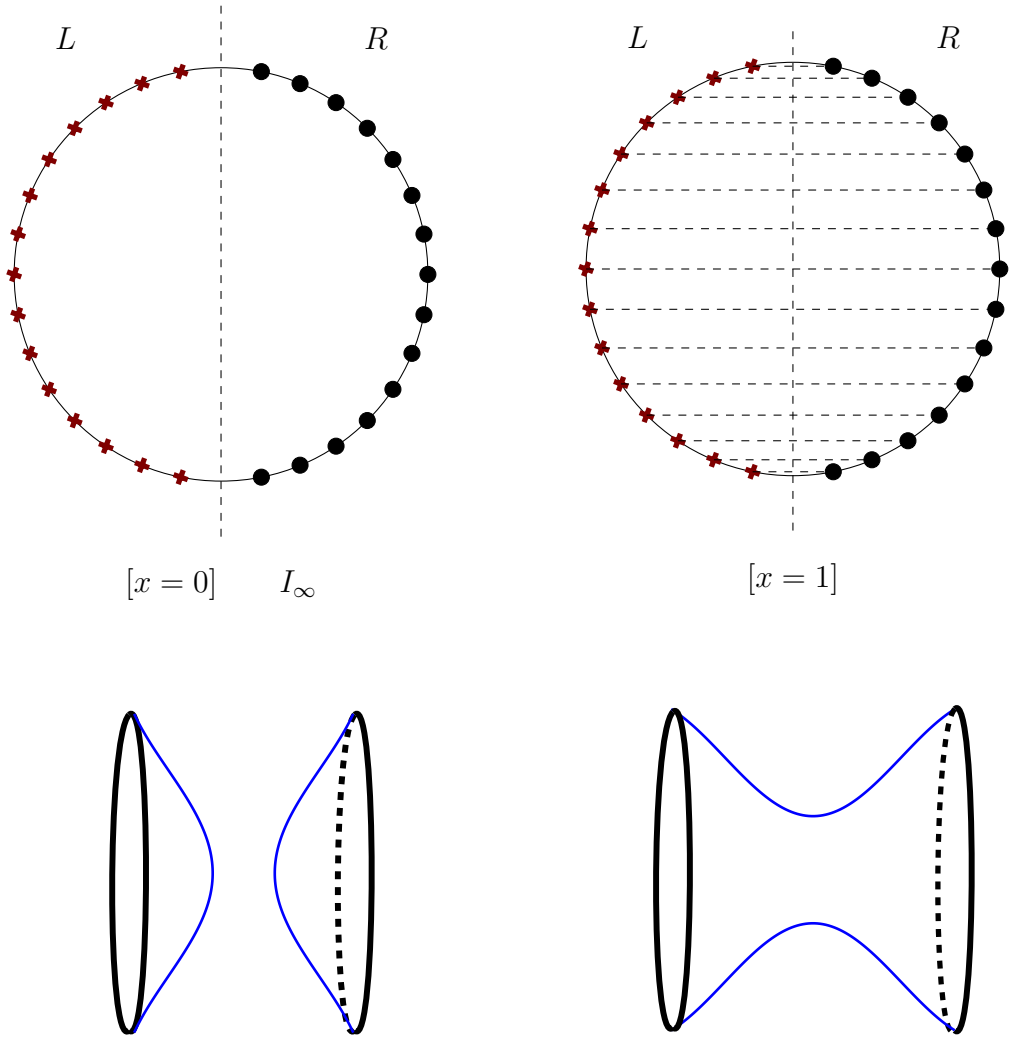
\begin{figure}
    \centering
    \begin{subfigure}[b]{0.45\textwidth}
        \centering
    \begin{tikzpicture}[scale=2.75]
        \draw[]circle(1) (0,0);
        \draw (-1,0) node[cross=4pt] {};
        \draw (-0.707,0.707) node[cross=4pt] {};
        \draw (-0.707,-0.707) node[cross=4pt] {};
        \draw (-0.382,0.923) node[cross=4pt] {};
        \draw (-0.382,-0.923) node[cross=4pt] {};
        \draw (-0.923,0.382) node[cross=4pt] {};
        \draw (-0.923,-0.382)  node[cross=4pt] {};
        \draw (-0.195,-0.981)  node[cross=4pt] {};
        \draw (-0.195,0.981)   node[cross=4pt] {};
        \draw (-0.981,-0.195)  node[cross=4pt] {};
        \draw (-0.981,0.195)   node[cross=4pt] {};
        \draw (-0.556,0.831)  node[cross=4pt] {};
        \draw (-0.556,-0.831)   node[cross=4pt] {};
        \draw (-0.831,-0.556)  node[cross=4pt] {};
        \draw (-0.831,0.556)    node[cross=4pt] {};
        \draw[dashed] (0,1.3)--(0,-1.3);
        \filldraw (1,0) circle(1.1pt) {};
        \filldraw (0.707,0.707) circle(1.1pt) {};
        \filldraw (0.707,-0.707) circle(1.1pt) {};
        \filldraw (0.382,0.923) circle(1.1pt) {};
        \filldraw (0.382,-0.923) circle(1.1pt) {};
        \filldraw (0.923,0.382) circle(1.1pt) {};
        \filldraw (0.923,-0.382) circle(1.1pt) {};
        \filldraw (0.195,-0.981)  circle(1.1pt) {};
        \filldraw (0.195,0.981)   circle(1.1pt) {};
        \filldraw (0.981,-0.195)  circle(1.1pt) {};
        \filldraw (0.981,0.195)  circle(1.1pt) {};
        \filldraw (0.556,0.831) circle(1.1pt) {};
        \filldraw (0.556,-0.831)  circle(1.1pt) {};
        \filldraw (0.831,-0.556)  circle(1.1pt) {};
        \filldraw (0.831,0.556)   circle(1.1pt) {};
        \draw[thick]  (-0.75,1.25)circle(0pt) node[anchor=north ] {$L$};
        \draw[thick]  (0.75,1.25)circle(0pt) node[anchor=north ] {$R$};
        \draw (0,-1.5) circle(0pt)node[ ] {$[x=0]\qquad I_\infty$};
    \end{tikzpicture}
    \newline
    \vspace{0.5cm}
    \newline
    \centering
    \begin{tikzpicture}[scale=.75]
        \draw[line width=2.25 pt] (-0.79*pi,2.8).. controls +(0.4, 0.4) and +(0.4, -0.4) .. (-0.79*pi,-2.8);
        \draw[line width=2.25 pt] (-0.79*pi,2.8).. controls +(-0.4, -0.4) and +(-0.4, -0.4) ..(-0.79*pi,-2.8);
        \draw[line width=2.25 pt] (4-0.79*pi,2.8).. controls +(0.4, 0.4) and +(0.4, -0.4) .. (4-0.79*pi,-2.8);
        \draw[line width=2.25 pt,dashed] (4-0.79*pi,2.8).. controls +(-0.4, -0.4) and +(-0.4, -0.4) ..(4-0.79*pi,-2.8);
        \draw[domain=-.9*pi:0.9*pi, samples=100, shift={(0,2)}, line width = 1, blue] plot ({0.9+.8*cos(deg(\x*0.85+pi))},\x-0.64*pi);
        \draw[domain=-.9*pi:0.9*pi, samples=100, shift={(0,2)}, line width = 1, blue] plot ({-1.8-.8*cos(deg(\x*0.85+pi))},\x-0.64*pi);
      \end{tikzpicture}
    \end{subfigure}
      \begin{subfigure}[b]{0.45\textwidth}
        \centering
        \begin{tikzpicture}[scale=2.75]
            \draw[dashed,opacity=0] (0,1.15)--(0,-1.3);
            \draw[]circle(1) (0,0);
            \draw (-1,0) node[cross=4pt] {};
            \draw (-0.707,0.707) node[cross=4pt] {};
            \draw (-0.707,-0.707) node[cross=4pt] {};
            \draw (-0.382,0.923) node[cross=4pt] {};
            \draw (-0.382,-0.923) node[cross=4pt] {};
            \draw (-0.923,0.382) node[cross=4pt] {};
            \draw (-0.923,-0.382)  node[cross=4pt] {};
            \draw (-0.195,-0.981)  node[cross=4pt] {};
            \draw (-0.195,0.981)   node[cross=4pt] {};
            \draw (-0.981,-0.195)  node[cross=4pt] {};
            \draw (-0.981,0.195)   node[cross=4pt] {};
            \draw (-0.556,0.831)  node[cross=4pt] {};
            \draw (-0.556,-0.831)   node[cross=4pt] {};
            \draw (-0.831,-0.556)  node[cross=4pt] {};
            \draw (-0.831,0.556)    node[cross=4pt] {};
            \draw[dashed] (-1,0) --(1,0) ;
            \draw[dashed] (-0.707,0.707) --(0.707,0.707) ;
            \draw[dashed] (-0.382,0.923) --(0.382,0.923) ;
            \draw[dashed] (-0.923,0.382)  --(0.923,0.382)  ;
            \draw[dashed] (-0.195,-0.981)  --(0.195,-0.981)  ;
            \draw[dashed] (-0.981,-0.195) --(0.981,-0.195) ;
            \draw[dashed] (-0.556,0.831)  --(0.556,0.831)  ;
            \draw[dashed] (-0.831,-0.556) --(0.831,-0.556) ;
            \draw[dashed] (-0.707,-0.707) --(0.707,-0.707) ;
            \draw[dashed] (-0.382,-0.923) --(0.382,-0.923) ;
            \draw[dashed] (-0.923,-0.382)  --(0.923,-0.382)  ;
            \draw[dashed] (-0.195,0.981)  --(0.195,0.981)  ;
            \draw[dashed] (-0.981,0.195) --(0.981,0.195) ;
            \draw[dashed] (-0.556,-0.831)  --(0.556,-0.831)  ;
            \draw[dashed] (-0.831,0.556) --(0.831,0.556) ;
            \filldraw (1,0) circle(1.1pt) {};
            \filldraw (0.707,0.707) circle(1.1pt) {};
            \filldraw (0.707,-0.707) circle(1.1pt) {};
            \filldraw (0.382,0.923) circle(1.1pt) {};
            \filldraw (0.382,-0.923) circle(1.1pt) {};
            \filldraw (0.923,0.382) circle(1.1pt) {};
            \filldraw (0.923,-0.382) circle(1.1pt) {};
            \filldraw (0.195,-0.981)  circle(1.1pt) {};
            \filldraw (0.195,0.981)   circle(1.1pt) {};
            \filldraw (0.981,-0.195)  circle(1.1pt) {};
            \filldraw (0.981,0.195)  circle(1.1pt) {};
            \filldraw (0.556,0.831) circle(1.1pt) {};
            \filldraw (0.556,-0.831)  circle(1.1pt) {};
            \filldraw (0.831,-0.556)  circle(1.1pt) {};
            \filldraw (0.831,0.556)   circle(1.1pt) {};
            \draw[thick]  (-.75,1.23)circle(0pt) node[anchor=north ] {$L$};
            \draw[thick]  (0.75,1.23)circle(0pt) node[anchor=north ] {$R$};
            \draw (0,-1.5) circle(0pt)node[ ] {$[x=1]$};
        \end{tikzpicture}
        \newline
        \vspace{0.5cm}
        \newline
        \centering
        \begin{tikzpicture}[scale=.75]
            \draw[domain=-.8*pi:0.8*pi, samples=100, shift={(0,2)}, line width = 1, blue] plot (\x, {1.05*cos(deg(\x+pi))});
            \draw[dashed,line width=2 pt] (0.8*pi,2.85).. controls +(-0.4, -0.4) and +(-0.4, -0.4) ..(0.8*pi,-2.85);
           \draw[line width=2.25 pt] (0.8*pi,2.85).. controls +(0.4, 0.4) and +(0.4, -0.4) .. (0.8*pi,-2.85);
          \draw[domain=-.8*pi:0.8*pi, samples=100, shift={(0,-2)}, line width = 1, blue] plot (\x, {1.05*cos(deg(\x))});
          \draw[line width=2.25 pt] (-0.79*pi,2.8).. controls +(0.4, 0.4) and +(0.4, -0.4) .. (-0.79*pi,-2.8);
          \draw[line width=2.25 pt] (-0.79*pi,2.8).. controls +(-0.4, -0.4) and +(-0.4, -0.4) ..(-0.79*pi,-2.8);
        \end{tikzpicture}
     \end{subfigure}
    \caption{Cartoon representing the type $I_{\infty}$ and type $II_1$ factors and the classical connectivity.}\label{fig:9} 
\end{figure} 

We will denote $B([\alpha])$ the corresponding von Neumann factor acting on $H([\alpha])$. In previous subsections we have considered the equivalence classes $[x=0]$ and $[x=1]$. Following \cite{EL1} we can say that the geometry $g([\alpha])$ is {\it connected} if $B([\alpha])$ is type $II$ or type $III$ and disconnected if it is type $I$. Generically we can define the analog of a TFD state in $H([\alpha])$ for $B([\alpha])$ type $I$ as a maximally entangled state in $H([\alpha])$. In the case $[x=0]$ corresponding to $I=I_L\cup I_R$ the so defined TFD state is  $|TFD\rangle=\sum_n c_n|n\rangle|\bar n\rangle$ for $|n\rangle$ and $\bar n\rangle$ a basis of the von Neumann Hilbert spaces $\otimes_{\alpha\in {I_L}}H_{\alpha}$ and of $\otimes_{\alpha\in {I_R}}H_{\alpha}$. By construction these states support a finite amount of quantum entanglement. The geometry $g([x=0])$ is {\it classically} disconnected although quantum effects can create different amounts of quantum entanglement.

Let us now consider the EPR equivalence class $[x=1]$. The algebraic relation between EPR and ER can be described interpreting the geometry $g([x=1])$ associated with the EPR equivalence class as representing an ER bridge. Equivalently the reference state of the EPR equivalence class $[x=1]$ will be interpreted as the ER (ER-bridge) state $|ER\rangle$. Note that the states $|TFD\rangle$ in $H([x=0])$ and $|ER\rangle$ in $H([x=1])$ are {\it orthogonal}.

Next we can use the fact that any hermitian operator $A$  in the type $II_1$ factor $R$ admits a {\it resolution of the identity} to define the associated {\it microcanonical states} as
\begin{equation}
|Micro(A)\rangle=:\int d(\lambda) d_R(E_{\lambda})|\lambda\rangle|\bar \lambda\rangle
\end{equation}
for $E_{\lambda}$ the spectral decomposition and for $d_R$ the type $II_1$ trace. Note that if $A$ defines the Hamiltonian the so defined state represents a type $II_1$ version of the microcanonical TFD state. Note that the key ingredient needed to define this microcanonical version is the existence of a spectral resolution of the hermitian operator representing the Hamiltonian. This is automatic in finite type $I_N$ and in finite type $II_1$.

For a given decomposition of $I$ and a given equivalence class $[\alpha]$ we can define the corresponding algebraic version of  the $L$ $R$ {\it entanglement wedges} as the spaces $B([\alpha])\Psi$ and $B([\alpha])^{'}\Psi$ for $\Psi$ the reference state used to define the equivalence class $[\alpha]$. We can again use as an algebraic measure of connectivity the coupling $\frac{d(B([\alpha])\Psi)}{d(B([\alpha])^{'}\Psi)}$ defined for the corresponding continuous dimension.
 
\end{appendices}
\section{Acknowledgements}
This work is supported through the grants CEX2020-001007-S and PID2021-123017NB-I00, funded by MCIN/AEI/10.13039/501100011033 and by ERDF A way of making Europe.


\begin{thebibliography}{99}
\bibitem{Hawking1}S. W. Hawking, "Breakdown of Predictability in Gravitational Collapse", Phys. Rev. D14,
2460–2473, 1976.
\bibitem{Hawking2} S. Hawking, "Particle Creation by Black Holes", Commun. Math. Phys. 43, 199–220, 1975.
[Erratum: Commun.Math.Phys. 46, 206 (1976)].
\bibitem{Bek1}
J.~D.~Bekenstein,``Black holes and entropy,''
Phys. Rev.D.7.2333
\bibitem{Mathur}
S. D. Mathur, "The Information paradox: A Pedagogical introduction", Class. Quant. Grav. 26,
224001, 2009, [arXiv:0909.1038 [hep-th]].
\bibitem{Bek2}
J. D. Bekenstein, "Black holes and the second law", Lett. Nuovo Cim. 4, 737–740, 1972.

\bibitem{RT}
S. Ryu and T. Takayanagi, "Holographic derivation of entanglement entropy from AdS/CFT",
Phys. Rev. Lett. 96, 181602, 2006, [arXiv:hep-th/0603001].

\bibitem{HRT} 
V. E. Hubeny, M. Rangamani and T. Takayanagi, "A Covariant holographic entanglement
entropy proposal", JHEP 07, 062, 2007, [arXiv:0705.0016 [hep-th]].
\bibitem{EW}
N. Engelhardt and A. C. Wall, "Quantum Extremal Surfaces: Holographic Entanglement
Entropy beyond the Classical Regime", JHEP 01, 073, 2015, [arXiv:1408.3203 [hep-th]].
\bibitem{QES1} 
G. Penington, "Entanglement Wedge Reconstruction and the Information Paradox", 2019,
[arXiv:1905.08255 [hep-th]].
\bibitem{QES2}
A. Almheiri, N. Engelhardt, D. Marolf and H. Maxfield, "The entropy of bulk quantum fields
and the entanglement wedge of an evaporating black hole", 2019, [arXiv:1905.08762 [hep-th]]
\bibitem{QES3}A. C. Wall, "Maximin Surfaces, and the Strong Subadditivity of the Covariant
Holographic Entanglement Entropy", Class. Quant. Grav. 31 (2014), no. 22 225007,
[arXiv:1211.3494].
\bibitem{QES4} C. Akers, N. Engelhardt, G. Penington, and M. Usatyuk, "Quantum Maximin Surfaces",
arXiv:1912.02799
\bibitem{Malda}
A.~Almheiri, T.~Hartman, J.~Maldacena, E.~Shaghoulian and A.~Tajdini,
``The entropy of Hawking radiation,''
Rev. Mod. Phys. \textbf{93} (2021) no.3, 035002
[arXiv:2006.06872 [hep-th]].
\bibitem{Gomez1}
C.~Gomez,
``The Algebraic Page Curve,''
[arXiv:2403.09165 [hep-th]].
\bibitem{Page}D. N. Page, "Information in black hole radiation", Phys. Rev. Lett. 71, 3743–3746, 1993,[hep-th/9306083]
\bibitem{Gia}
G.~Dvali and C.~Gomez,
``Black Hole's Quantum N-Portrait,''
Fortsch. Phys. \textbf{61} (2013), 742-767
doi:10.1002/prop.201300001
[arXiv:1112.3359 [hep-th]].
\bibitem{MvN1}F. J. Murray and J. v. Neumann, "On Rings of Operators"
Annals of Mathematics , Jan., 1936, Second Series, Vol. 37, No. 1 (Jan., 1936),
pp. 116-229
\bibitem{Jones}
V.F.R. Jones "Von Neumann Algebras" Vanderbilt 2015
\bibitem{Popa}
C. Anantharaman and
S. Popa, "An introduction to II1 factors"

\bibitem{Araki}
H.Araki,
"Type of von Neumann Algebra Associated with Free Field"
Progress of Theoretical Physics, Vol. 32, No. 6, December 1964
\bibitem{Liu1}
 S. Leutheusser and H. Liu, "Emergent times in holographic duality" 2112.12156.
 \bibitem{Liu2}
 S. Leutheusser and H. Liu, [arXiv:2112.12156 [hep-th]].
 \bibitem{Witten0}
E.~Witten,
"Gravity and the Crossed Product''
[arXiv:2112.12828 [hep-th]].
\bibitem{Witten1} 
V.~Chandrasekaran, R.~Longo, G.~Penington and E.~Witten,
``An Algebra of Observables for de Sitter Space,''
[arXiv:2206.10780 [hep-th]].
\bibitem{Witten2}
V.~Chandrasekaran, G.~Penington and E.~Witten,
``Large N algebras and generalized entropy,''
[arXiv:2209.10454 [hep-th]].
\bibitem{Witten3}
E.~Witten,
``A background-independent algebra in quantum gravity,''
JHEP \textbf{03} (2024), 077
[arXiv:2308.03663 [hep-th]].
\bibitem{EPR}
J.~Maldacena and L.~Susskind,
``Cool horizons for entangled black holes,''
Fortsch. Phys. \textbf{61} (2013), 781-811
doi:10.1002/prop.201300020
[arXiv:1306.0533 [hep-th]].

\bibitem{Dopli}
Doplicher, S, Kaskler, D.  Robinson, D., "Covariance algebras in field theory and
statistical mechanics". Comm. Math. Phys., 3 (1966), 1-28.
\bibitem{Takesaki} M.Takesaki
"Duality for Crossed Products and the Structure
of von  Neumann Algebras of type III" Berkeley preprint 1973
\bibitem{SvN}
Mackey, G. W., "A theorem of Stone and von Neumann." Duke Math. J., 16 (1949), 313-326.
\bibitem{CT}
A. Connes and M. Takesaki,
The Flow of Weights on Factors of Tyoe III
Tohoku Math. Journal.
29 (1977), 473-575.
\bibitem{Gomez2}
C.~Gomez,
``Traces and Time: a de Sitter Black Hole correspondence,''
[arXiv:2307.01841 [hep-th]].
\bibitem{Gomez3}
C.~Gomez,
``On the algebraic meaning of quantum gravity for closed Universes,''
[arXiv:2311.01952 [hep-th]].
\bibitem{Page1}
D. N. Page, Phys. Rev. Lett. 71, 1291 (1993), gr-qc/9305007.

\bibitem{Longo}
R.~Longo and E.~Witten,
``A note on continuous entropy,''
Pure Appl. Math. Quart. \textbf{19} (2023) no.5, 2501-2523
[arXiv:2202.03357 [math-ph]].
\bibitem{MvN2}F. J. Murray and J. von Neumann,
"On Rings of Operators. IV"
Annals of Mathematics , Oct., 1943, Second Series, Vol. 44, No. 4 (Oct., 1943), pp.716-808
\bibitem{EL1} 
N.~Engelhardt and H.~Liu,
``Algebraic ER=EPR and Complexity Transfer,''
[arXiv:2311.04281 [hep-th]].
\bibitem{Gomez4}
C.~Gomez,
``Clocks, Algebras and Cosmology,''
[arXiv:2304.11845 [hep-th]].
\bibitem{vN1}J. Von Neumann,
"On infinite direct products"
Compositio Mathematica, tome 6 (1939), p. 1-77
\bibitem{AW} H. Araki and E. J. Woods
"A Classification of Factors"
Publ. RIMS, Kyoto Univ. Ser. A
Vol. 3 (1968), pp. 51—130
\bibitem{Powers}
Powers, R. T., "Representations of uniformly hyperfinite algebras and their
associated von Neumann rings", Ann. of Math. 86 (1967), 138-171.
\bibitem{haagerup} U.Haagerup
"Connes' bicentralizer problem and uniqueness of
the injective factor of type $III_1$ 1985

\end{thebibliography}
\end{document}